\documentclass[12pt]{iopart}
\usepackage[utf8]{inputenc}
\usepackage[colorlinks=true,linkcolor=blue,citecolor=blue,urlcolor=blue]{hyperref}
\usepackage[T1]{fontenc}
\usepackage{bm}
\usepackage{subcaption}
\usepackage{multirow}
\usepackage{float}
\usepackage{setspace}
\usepackage{graphicx}
\usepackage{siunitx}
\DeclareSIUnit\rydberg{Ry}
\usepackage{booktabs}
\usepackage{balance} 
\usepackage{cite}
\usepackage{lineno}
\usepackage{ulem}
\usepackage{titlesec}
\usepackage[compatibility=false]{caption}

\usepackage{booktabs}  
\usepackage{tabularx}  
\usepackage{array}      
\usepackage[T1]{fontenc}
\usepackage[utf8]{inputenc}
\usepackage{lmodern}               % clean default font
\usepackage{geometry}
\titleformat{\subsubsection}[block]
  {\normalfont\normalsize\itshape}
 {\thesubsubsection}
  {1em}{}

\titlespacing*{\subsubsection}{0pt}{3.25ex plus 1ex minus .2ex}{1.5ex plus .2ex}

\begin{document}

\title[Chiral spin-textures in van der Waals heterostructures]{Chiral spin-textures in van der Waals heterostructures}
\author{Nihad Abuawwad$^{1}$, Samir Lounis$^{2}$}
\address{$^1$ Peter Gr\"unberg Institut, Forschungszentrum J\"ulich \& JARA, 52425 J\"ulich, Germany}
\address{$^2$ Institute of Physics and Halle-Berlin-Regensburg Cluster of Excellence CCE, Martin-Luther-University Halle-Wittenberg, 06099 Halle, Germany}
\eads{\mailto{n.abuawwad@fz-juelich.de}, \mailto{samir.lounis@physik.uni-halle.de}}
%%%%%%%%%%%%%%%%%%%%%%%%%%%%%%%%%%%%%%%%%%%%%%%%%%%%%%%%%%%%%

\begin{abstract}

Chiral spin textures such as skyrmions have attracted considerable attention due to their nontrivial topology, chirality,  stability at the nanoscale, and potential for low-power spintronic devices. The recent discovery of intrinsic magnetism in van der Waals (vdW) materials and the ability to engineer their heterostructures has opened a new platform to study and manipulate such textures. In these layered systems, atomically sharp interfaces, strong spin–orbit coupling, and tunable symmetry breaking provide unique opportunities to stabilize and control chiral magnetic states. This review summarizes the fundamental mechanisms underlying the formation of chiral spin textures in vdW heterostructures, including the roles of exchange interactions, magnetic anisotropy, Dzyaloshinskii–Moriya interaction, and dipolar effects. We highlight key experimental advances in the observation and manipulation of chiral textures, discuss their dynamical properties and transport signatures, while overviewing selected theoretical investigations. Finally, we outline current challenges and future directions toward realizing robust, room-temperature chiral spin textures for practical spintronic technologies.

\end{abstract}
%%%%%%%%%%%%%%%%%%%%%%%%%%%%%%%%%%%%%%%%%%%%%%%%%%%%%%%%%%%%%

\maketitle

\section{Introduction}

The study of two-dimensional (2D) materials has revealed remarkable physical phenomena that often defy conventional expectations. One particularly intriguing aspect is magnetism in reduced dimensionality. In 1966, it was proven that 2D systems with continuous spin-rotational symmetry cannot sustain long-range magnetic order at finite temperature~\cite{mermin}. This implies that magnetic order in 2D relies on interactions that break spin-rotational invariance, such as external magnetic fields or magnetocrystalline anisotropy. These terms are typically much weaker than exchange interactions and would therefore suggest vanishingly low ordering temperatures. 
It was thus surprising when intrinsic 2D magnets were experimentally discovered, namely the antiferromagnet FePS$_3$, which revealed that magnetic order in the 2D limit can also arise from anisotropic exchange interactions~\cite{Lee2016,Wildes2017}. Shortly thereafter, the ferromagnets CrI$_3$ and Cr$_2$Ge$_2$Te$_6$ were identified, demonstrating that sufficiently strong anisotropy can stabilize long-range magnetic order even in atomically thin layers~\cite{Huang2017,Gong2017}. Since these discoveries, rapid progress has been made in expanding the family of 2D magnets and uncovering their diverse magnetic and topological properties~\cite{Gibertini2019,doi:10.1126/science.aav4450, 10.1063/5.0023729, Sierra2021}.

The van der Waals (vdW) nature of these materials enables stacking into heterostructures with atomically sharp interfaces and minimal lattice mismatch. Such vdW heterostructuring provides opportunities to engineer quantum materials with tailored functionalities, including proximity-induced effects, enhanced spin–orbit coupling (SOC), and inversion-symmetry breaking~\cite{novoselov2016van,geim2021van}. The combination of strong SOC and broken inversion symmetry gives rise to the Dzyaloshinskii–Moriya interaction (DMI), originally proposed by Dzyaloshinskii and formulated microscopically by Moriya~\cite{DZYALOSHINSKY1958241,Moriya1960}, which is the key microscopic mechanism responsible for stabilizing chiral magnetism. This interaction favors noncollinear spin configurations, leading to spin textures such as helices, skyrmions, and merons that are otherwise absent in simple collinear ferromagnets and antiferromagnets. In addition to these intrinsic advantages, vdW systems offer multiple external control knobs that further reinforce the stabilization and manipulation of chiral spin textures. 

Proximity to heavy transition-metal dichalcogenides or metallic layers can strongly enhance SOC and induce DMI~\cite{hung2023enhanced,rosenberger2024proximity,che2024magnetic, Liang2020, Kumar2020APL, Shi2019NatNano, Wu2020NatCommun,Zollner2025}, while electrostatic gating and chemical doping can tune carrier density and magnetic anisotropy, enabling phase transitions between ferromagnetic, antiferromagnetic, and noncollinear states~\cite{zheng2020gate}. Strain engineering, achievable through substrate choice or mechanical bending of thin flakes, directly modifies exchange, anisotropy, and DMI, thereby stabilizing different types of skyrmions~\cite{Zhou2025, Ren2023, Abuawwad2023PRB}. Twist-angle control in bilayer vdW magnets further introduces moiré superlattices, where spatially modulated interactions can lead to periodic skyrmion or meron crystals~\cite{tong2018moire,akram2020twisted}. 

Beyond static approaches, ultrafast optical techniques provide dynamic pathways to manipulate chiral textures through photoinduced anisotropy or Floquet engineering of magnetic interactions~\cite{jiang2020optical,kimel2019ultrafast}. Electric-field control of magnetism, enabled by magnetoelectric coupling or multiferroic proximity effects, has also been predicted as a promising route toward low-power, reversible switching of skyrmions~\cite{Abuawwad2024,huang2022ferroelectric,liu2024multiferroic}. Altogether, these intrinsic and extrinsic mechanisms highlight the versatility of vdW heterostructures for stabilizing, tuning, and dynamically manipulating chiral spin textures.

Beyond static imaging, transport properties provide unique and versatile signatures of chiral magnetism in reduced dimensions. The topological Hall effect (THE), originating from the emergent electromagnetic field associated with real-space Berry curvature of skyrmions, has been widely employed to identify chiral textures in both bulk and thin-film magnets, and has recently been reported in vdW magnets such as Fe$_3$GeTe$_2$ and related heterostructures~\cite{Wu2020,ding2020observation,wang2020direct}. In addition to THE, the anomalous Hall effect (AHE) remains a sensitive probe of magnetization and has been extensively studied in vdW ferromagnets, where it is strongly influenced by spin–orbit coupling and Berry curvature effects~\cite{liu2018anomalous,kim2018large}. More recently, the so-called noncollinear Hall effect was predicted in 2D magnets, demonstrating that even in the absence of net magnetization, real-space chirality and SOC can generate transverse transport signals~\cite{bouaziz2021transverse}. Complementary measurements such as spin Hall magnetoresistance (SMR) provide a direct probe of interfacial spin–orbit torques and their coupling to noncollinear magnetism, with room-temperature SMR already reported in vdW flakes~\cite{feringa2022smr}. 

Other transport-based probes have also proven essential. Tunneling magnetoresistance (TMR) in magnetic tunnel junctions based on vdW magnets, such as CrI$_3$ and Fe$_3$GeTe$_2$, has demonstrated large spin-filtering effects and interlayer-dependent tunneling conductance~\cite{jiang2018interlayer,klein2018probing}. Non-local spin valve geometries using graphene and vdW ferromagnets have been developed to detect spin currents and magnetization switching in atomically thin devices~\cite{ghosh2020proximity,ghosh2021spin}. In parallel, magneto-optical Kerr effect (MOKE) measurements have provided sensitive, noninvasive detection of chiral domain walls and skyrmion dynamics in vdW ferromagnets~\cite{jiang2020optical}. MOKE thus serves as a robust fingerprint of chiral spin textures in vdW systems and highlights their potential for nonvolatile information storage, electrical readout, and spintronic device integration.

Altogether, these developments place vdW heterostructures at the forefront of research on chiral magnetism. The combination of low dimensionality, strong SOC, proximity-induced effects, and topological textures offers a versatile platform for exploring novel physics and for realizing next-generation spintronic and magnonic devices. In the following sections, we review the fundamental mechanisms underlying chiral spin textures in vdW materials, survey key experimental advances, examine their dynamical and transport properties, and highlight selected insights from recent simulations. This review is not intended to be exhaustive; instead, we focus on a representative selection of recent developments. 

\begin{figure}[H]
    \centering
    \includegraphics[width=\textwidth]{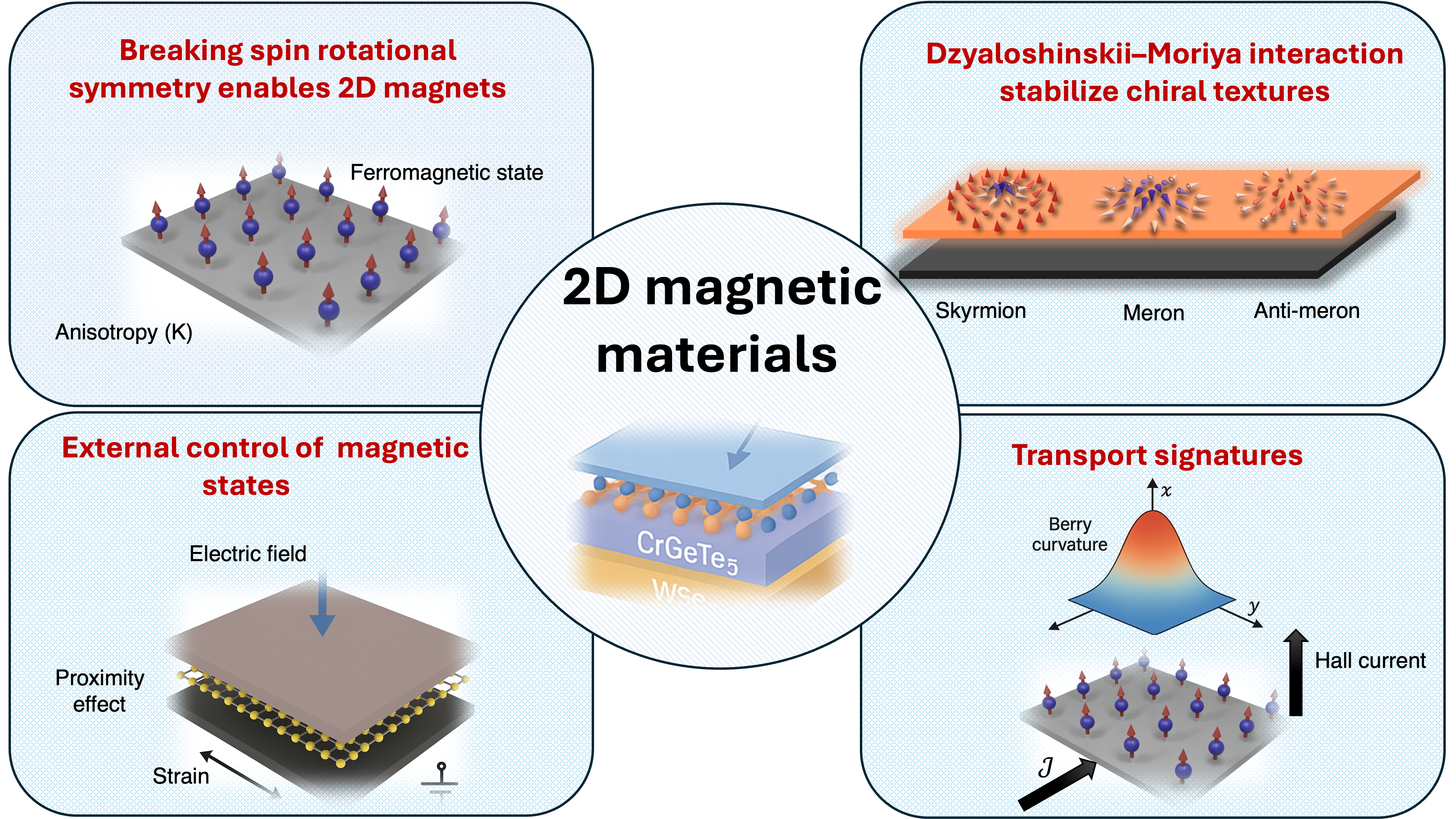}
    \caption{
        Schematic overview of key concepts in 2D magnetic materials.
        (Top left) Breaking spin rotational symmetry via magnetic anisotropy enables two-dimensional ferromagnetism. 
        (Top right) The Dzyaloshinskii–Moriya interaction (DMI) stabilizes chiral magnetic textures such as skyrmions, merons, and antimerons. 
        (Bottom left) External control of magnetic states can be achieved through electric fields, strain, or proximity effects. 
        (Bottom right) Nontrivial Berry curvature in these systems leads to transport signatures such as the anomalous Hall current.
    }
    \label{fig:2Dmagneticsummary}
\end{figure}

\section{Fundamental concepts}

\subsection{Fundamental magnetic interactions}\label{sec:fundamental-interactions}
Magnetism in solids arises from the interplay of several microscopic interactions that determine how atomic spins align and evolve under external influences. These interactions can be described at two complementary levels: the \textit{atomistic} description, which treats spins as discrete magnetic moments localized at atomic sites, and the \textit{micromagnetic} description, which represents the magnetization as a continuous vector field varying smoothly in space (See Table.\ref{tab:magnetic_interactions}).  
Together, these frameworks capture the essential physics behind magnetic phenomena such as ferromagnetism, antiferromagnetism, spin spirals, and topologically nontrivial textures like skyrmions and merons. 

\begin{table*}[htbp]
\centering
\caption{Magnetic interactions in both their atomistic and micromagnetic formulations.}
\label{tab:magnetic_interactions}
\footnotesize
\setlength{\tabcolsep}{6pt}
\renewcommand{\arraystretch}{1.3}

\begin{tabularx}{\textwidth}{@{} >{\raggedright\arraybackslash}X @{}}
\toprule
\textbf{Magnetic interactions} \\
\midrule

\textbf{Exchange interaction}\\
Quantum-mechanical interaction aligning neighboring spins. Positive (negative) $J$ favors ferromagnetism (antiferromagnetism). Competing interactions can produce frustration and spin spirals.

\medskip
\textbf{Atomistic:}

{\centering
$E_{\mathrm{ex}} = - \sum_{\langle i,j \rangle} J_{ij} (\boldsymbol{\mu}_i \cdot \boldsymbol{\mu}_j)$,\par} where $\boldsymbol{\mu}$ indicates an atomic spin moment.

\textbf{Micromagnetic:}

{\centering
$E_{\mathrm{ex}} = A \int (\nabla \mathbf{m})^2 \, dV$, \par}
where $A$ is the exchange constant, $\mathbf{m}=\mathbf{M}/M_s$ is the normalized magnetization with saturation magnetization $M_s$, and the integral is taken over the sample volume.

\\
\midrule

\textbf{Dzyaloshinskii--Moriya interaction (DMI)}\\
Antisymmetric exchange in non-centrosymmetric systems with strong spin--orbit coupling. It generates chirality and stabilizes spin spirals and skyrmions.

\medskip
\textbf{Atomistic:}

{\centering
$E_{\mathrm{DMI}} = \sum_{\langle i,j \rangle} \mathbf{D}_{ij} \cdot (\boldsymbol{\mu}_i \times \boldsymbol{\mu}_j)$
\par}

\textbf{Micromagnetic:}

{\centering
$E_{\mathrm{DMI}} = D \int \mathbf{m} \cdot (\nabla \times \mathbf{m}) \, dV$
\par}

\\
\midrule

\textbf{Magnetic anisotropy}\\
Originates from spin--orbit coupling and defines easy and hard magnetization axes.

\medskip
\textbf{Atomistic:}

{\centering
$E_{\mathrm{ani}} = - \sum_i K_i (\boldsymbol{\mu}_i \cdot \mathbf{e}_i)^2$
\par}

\textbf{Micromagnetic:}

{\centering
$E_{\mathrm{ani}} = K \int \left(1 - (\mathbf{m} \cdot \mathbf{e})^2 \right) dV$
\par}

\\
\midrule

\textbf{Zeeman interaction}\\
Coupling of magnetic moments to an external magnetic field.

\medskip
\textbf{Atomistic:}

{\centering
$E_Z = - \sum_i \boldsymbol{\mu}_i \cdot \mathbf{B}$
\par}

\textbf{Micromagnetic:}

{\centering
$E_Z = - \mu_0 \int \mathbf{M} \cdot \mathbf{H}_{\mathrm{ext}} \, dV$
\par}

\\
\midrule

\textbf{Dipolar interaction}\\
Long-range interaction due to stray magnetic fields controlling domains and textures.

\medskip
\textbf{Atomistic:}

{\centering
$E_{\mathrm{dip}} = \frac{\mu_0}{4\pi} \sum_{i<j} \left[
\frac{\boldsymbol{\mu}_i \cdot \boldsymbol{\mu}_j}{r_{ij}^3}
- \frac{3(\boldsymbol{\mu}_i \cdot \mathbf{r}_{ij})(\boldsymbol{\mu}_j \cdot \mathbf{r}_{ij})}{r_{ij}^5}
\right]$
\par}

\textbf{Micromagnetic:}

{\centering
$E_{\mathrm{dip}} = \frac{\mu_0}{2} \int |\mathbf{H}_d|^2 \, dV$
\par}

\\
\bottomrule
\end{tabularx}

\end{table*}

\subsection{First-principles evaluation of magnetic interactions}

Magnetic interactions in solids can be extracted from first principles using several complementary approaches that differ in both their theoretical formulation and computational implementation.
A first class consists of total-energy mapping methods, in which the energies of a set of magnetic configurations are computed self-consistently and mapped onto an effective spin Hamiltonian, such as the Heisenberg model~\cite{Whangbo2003,Lee2012}. Within this framework, specialized schemes such as the four-state method enable the extraction of individual components of the exchange tensor and Dzyaloshinskii--Moriya interaction (DMI) by constructing carefully chosen spin configurations~\cite{Xiang2011,Sabani2020}.
A second class comprises reciprocal-space approaches, including frozen-magnon and spin-spiral methods, where the magnetic energy dispersion $E(\mathbf{q})$ is calculated and subsequently Fourier transformed or fitted to obtain exchange interactions~\cite{Halilov1998}. These approaches are commonly formulated using the generalized Bloch theorem~\cite{Sandratskii1986}.
A third class involves linear-response methods, in which magnetic interactions are derived from the response of the system to infinitesimal spin perturbations. This is typically formulated in terms of transverse magnetic susceptibilities within time-dependent density functional theory (TDDFT) or many-body perturbation theory, allowing direct access to exchange parameters and magnon spectra~\cite{Sasioglu2010,Lounis2010,Buczek2011,Lounis2011,Schweflinghaus2014,Lounis2015,TancogneDejean2020,Bouaziz2020,Olsen2021,dosSantos2026}.
An important and widely used approximation is provided by the magnetic force theorem (MFT), which evaluates magnetic energy differences within a frozen-potential approximation. In Green-function-based methods, such as Korringa--Kohn--Rostoker (KKR) approaches, this leads to the well-known Liechtenstein--Katsnelson--Antropov--Gubanov (LKAG) formalism for infinitesimal spin rotations and direct computation of exchange parameters~\cite{Liechtenstein1987,Udvardi2003,Ebert2011,PhysRevB.83.024401,Szilva2023}. In full-potential linearized augmented plane-wave (FLAPW) methods, such as those implemented in FLEUR, the force theorem is typically employed in a non--Green-function form, for example, through spin-spiral or small-rotation calculations~\cite{Kurz2004,Zimmermann2019}. In addition, recent developments allow the application of Green-function-like MFT formulations within FLAPW frameworks via Wannier-based or tight-binding post-processing approaches, thereby bridging the gap between Green-function and plane-wave methodologies~\cite{Tb2j}. The validity and limitations of the force theorem have been discussed extensively in Refs.~\cite{Bruno2003,LounisPRB2010,Solovyev2021}.
Finally, magnetic interactions can also be obtained through effective-model construction, where first-principles electronic structure calculations are downfolded to low-energy Hamiltonians, such as Wannier-based tight-binding or Hubbard models. In this framework, magnetic couplings are derived from superexchange or related mechanisms, tracing back to Anderson’s theory~\cite{Mazurenko2016,Ramadan2021,Anderson1950}.
Together, these approaches form a hierarchy ranging from fully self-consistent total-energy methods to linear-response and force-theorem-based techniques, as well as low-energy model constructions for correlated systems. An overview of commonly used DFT codes and their corresponding methodologies is summarized in Table~\ref{tab:dft_codes}.

\begin{table*}[htb!]
\centering
\caption{Overview and examples of DFT codes and methods used to extract magnetic interactions.}
\label{tab:dft_codes}
\footnotesize
\setlength{\tabcolsep}{4pt}
\renewcommand{\arraystretch}{1.15}

\begin{tabularx}{\textwidth}{@{} l l >{\raggedright\arraybackslash}X @{}}
\toprule
\textbf{Code} & \textbf{Formalism} & \textbf{Methods} \\
\midrule

JuKKR~\cite{jukkrcode} & KKR (multiple scattering) & 
Magnetic force theorem (Green function / LKAG) \\

FLEUR~\cite{fleurcode} & Full-potential LAPW & 
Spin spirals (generalized Bloch), Total-energy mapping, Green function \\

ELK~\cite{elkcode} & Full-potential LAPW & 
Spin spirals, Total-energy mapping \\

WIEN2k~\cite{wien2kcode} & Full-potential LAPW & 
Total-energy mapping \\

VASP~\cite{vaspcode} & Plane-wave pseudopotential & 
Total-energy mapping, spin spirals \\

Quantum ESPRESSO~\cite{qecode} & Plane-wave pseudopotential & 
Total-energy mapping \\

OpenMX/SIESTA~\cite{openmxcode,siestacode} & Local orbital DFT & 
Total-energy mapping \\

\bottomrule
\end{tabularx}
\end{table*}
\newpage
In addition to the fundamental bilinear interactions, the same first-principles frameworks can be systematically extended to evaluate higher-order magnetic interactions, such as biquadratic exchange, four-spin, and chiral multi-spin terms~\cite{Brinker2019,Brinker2020,Hoffmann2020,Grytsiuk2020,Laszloffy2019}. In total-energy mapping approaches, this requires going beyond pairwise spin rotations and constructing sets of noncollinear magnetic configurations that probe higher-order angular dependencies~\cite{Kurz2004,Antal2008}. For instance, the biquadratic interaction $B_{ij}$ between two sites $i$ and $j$ can be extracted by computing the total energy as a function of the relative angle between two spins and fitting it to an extended form $E_{ij} = -J_{ij}\cos\theta_{ij} - B_{ij}\cos^2\theta_{ij}$, where deviations from a purely Heisenberg-like cosine behavior directly quantify the higher-order contribution~\cite{Eckardt2020Biquadratic}. In periodic systems, spin-spiral calculations can be used by correlating deviations in the dispersion $E(\mathbf{q})$ signal to the presence of non-Heisenberg interactions~\cite{Sandratskii1998}. 
For multi-spin interactions, such the four-spin isotropic term, larger supercells and carefully designed magnetic configurations are required. One of the key ideas is to construct sets of spin configurations in which bilinear contributions cancel by symmetry, allowing the residual energy differences to be uniquely mapped onto multi-spin terms~\cite{Hoffmann2020}. Similarly, chiral higher-order interactions can be obtained by comparing configurations with opposite scalar spin chirality, which necessitates fully noncollinear calculations including spin--orbit coupling~\cite{Brinker2020}. Within Green-function-based implementations of the magnetic force theorem, higher-order interactions can be accessed by extending the infinitesimal rotation formalism to higher-order derivatives of the band energy with respect to spin rotations, providing an efficient alternative to supercell-based approaches~\cite{Brinker2019,Brinker2020,LounisPRB2010,Lounis2020}.

\newpage
\subsection{Topological Magnetism}\label{sec3-7}

Topological magnetism refers to magnetic states whose stability arises not only from energetics but also from their topology. These topologically protected spin textures cannot be continuously transformed into trivial ferromagnetic or paramagnetic states without crossing an energy barrier or introducing singularities. Their unique stability and particle-like nature make them central to modern spintronics and magnetism research.

\subsubsection{Domain walls and spiral states}

\begin{figure}[H]
    \centering
    \includegraphics[width=\linewidth]{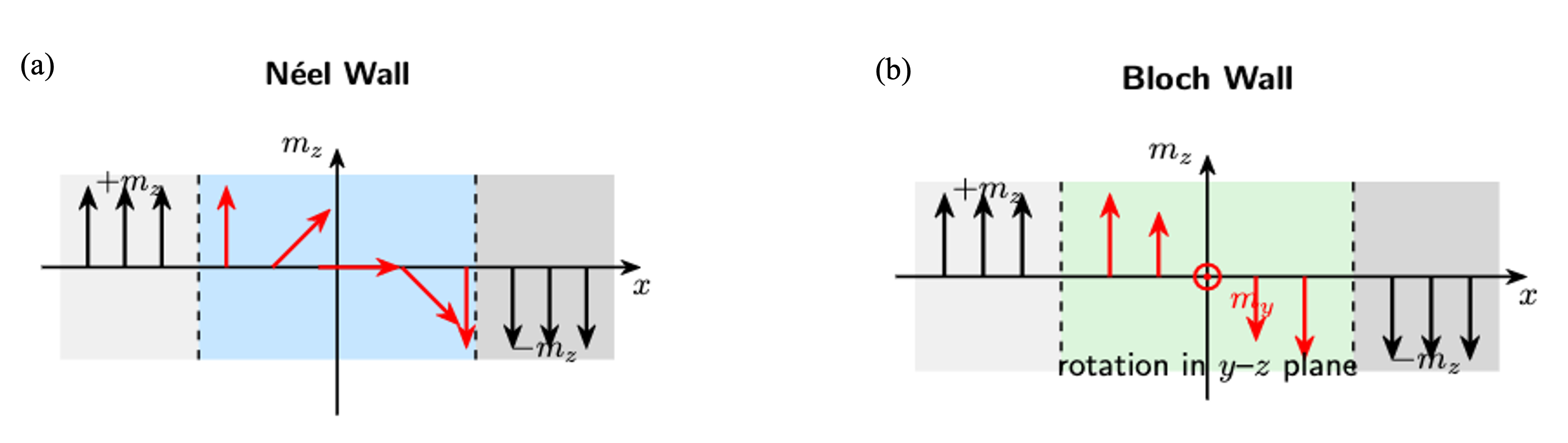}
    \caption{
Schematic illustration of chiral domain walls in a ferromagnet with broken inversion symmetry.  
(a) Chiral Néel wall: magnetization rotates within the $x$--$z$ plane, producing an in-plane component along $x$.  
(b) Bloch wall: rotation occurs in the $y$--$z$ plane, leading to an in-plane component along $y$.  
The fixed sense of rotation is governed by the symmetry of the Dzyaloshinskii--Moriya interaction (DMI).}
\label{fig:domain_walls}
\end{figure}

Before discussing truly topological spin textures, it is essential to consider their one-dimensional precursors: domain walls and spiral states.  
In ferromagnets with broken inversion symmetry, the competition between exchange interaction, and the Dzyaloshinskii--Moriya interaction leads to twisted magnetic configurations.  
In systems with DMI, it induces a fixed sense of rotation within domain walls, producing \textit{chiral Néel} or \textit{Bloch walls}, depending on the DMI symmetry (see Fig.~\ref{fig:domain_walls}).  
Extending these chiral twists to two dimensions gives rise to \textit{spin spirals}, which are periodic modulations of the magnetization vector in space.  
 The spiral period is determined by the balance between the DMI strength and the exchange stiffness, \( \lambda \sim J/D \), and can transition to more localized textures under confinement or external fields.  
Such spiral and domain-wall states form the foundation for higher-dimensional topological structures such as skyrmions and merons.

\newpage
\subsubsection{Skyrmions and Merons}

\begin{figure}[H]
    \centering
    \includegraphics[width=\textwidth]{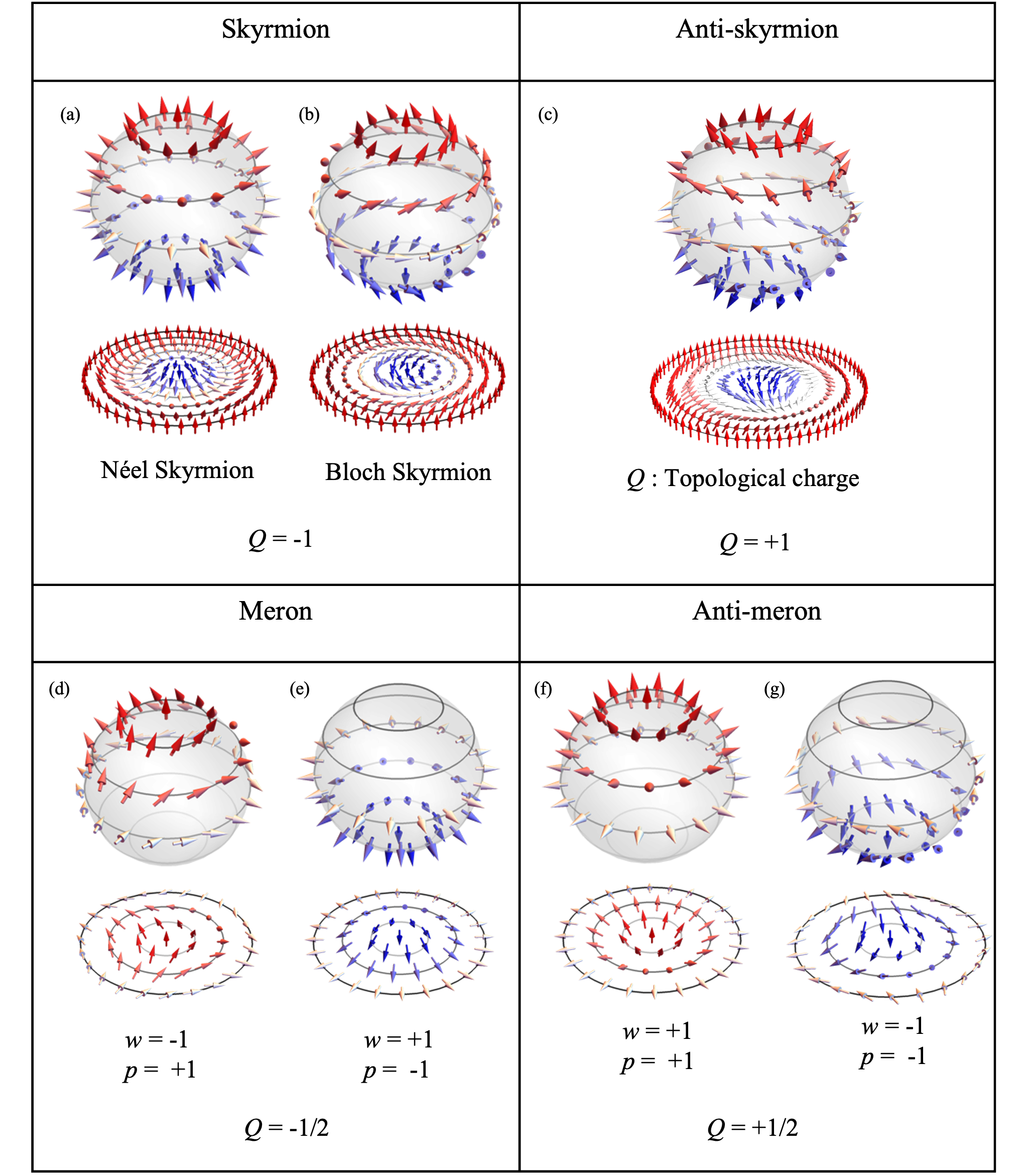}
    \caption{
        Spin textures of various magnetic topological objects with their projection onto the unit sphere.  
        (a,b) Néel- and Bloch-type skyrmions (\(Q=-1\));  
        (c) Anti-skyrmion (\(Q=+1\));  
        (d,e) Merons (\(Q=-1/2\));  
        (f,g) Anti-merons (\(Q=+1/2\)).  
        Here \(p=\pm 1\) denotes polarity (core orientation) and \(w=\pm 1\) denotes vorticity (sense of in-plane rotation).
    }
    \label{figure3_5}
\end{figure}

Magnetic skyrmions are two-dimensionally confined~\ref{figure3_5}, particle-like spin textures stabilized by the interplay between symmetric exchange and antisymmetric DMI. Their type: Néel, Bloch, or anti-skyrmion is dictated by the crystal symmetry governing the DMI form.  
In noncentrosymmetric bulk magnets with cubic symmetry, such as MnSi and FeGe, \textit{bulk DMI} stabilizes Bloch-type skyrmions, characterized by tangential spin rotation around the core~\cite{Muhlbauer2009,Yu2011}. At interfaces with broken inversion symmetry, \textit{interfacial DMI} favors Néel-type skyrmions, where spins rotate radially~\cite{Heinze2011,Emori2013,Chen2013,Wu2020}. In systems of reduced symmetry, \textit{anisotropic DMI} gives rise to \textit{anti-skyrmions} with nonuniform in-plane rotation and a reversed topological charge \(Q=+1\)~\cite{Nayak2017}.  

The topology of a skyrmion is quantified by its \textit{topological charge} \(Q\), defined as:
\begin{equation}
Q = \frac{1}{4\pi} \int d^2r \, \mathbf{m} \cdot \left( \frac{\partial \mathbf{m}}{\partial x} \times \frac{\partial \mathbf{m}}{\partial y} \right),
\end{equation}
where \( \mathbf{m} \) is the unit vector of local magnetization.  
In spherical coordinates, this can be expressed as
\begin{equation}
\mathbf{m}(\mathbf{r}) = ( \cos\Phi(\phi)\sin\theta(r),  \sin\Phi(\phi)\sin\theta(r),  \cos\theta(r) )
\end{equation}
yielding the compact relation
\begin{equation}
Q = p \cdot w,
\end{equation}
where polarity \(p=\pm 1\) indicates whether the core magnetization points up or down, and vorticity \(w=\pm 1\) defines the sense of in-plane rotation. Typically, Bloch and Néel skyrmions have \(Q=-1\), while anti-skyrmions possess \(Q=+1\)~\cite{Bogdanov2001,Nagaosa2013}.  

Beyond integer-charged skyrmions, \textit{merons} and \textit{anti-merons} represent fractional topological defects with half-integer topological charge: \(Q=-1/2\) for merons and \(Q=+1/2\) for anti-merons~\cite{Yu2018}. 
Unlike skyrmions, which cover the full spin sphere, merons wrap only half of it, corresponding to a hemispherical mapping. Their stabilization requires a delicate balance between DMI, magnetic anisotropy, and dipolar interactions. In particular, a strong in-plane anisotropy combined with an out-of-plane DMI component (\(D_z\)) can cant spins such that only half of the spin sphere is covered, giving rise to these \textit{half-skyrmion} states~\cite{Gao2019, Abuawwad2023PRB, Abuawwad2024 }.   
A meron–antimeron pair can recombine to form a complete skyrmion, underscoring their role as the fundamental building blocks of topological magnetism.

\section{Experimental observations of chiral spin textures in 2D-vdW materials}

Recent advances in 2D magnetism have led to the experimental realization of chiral spin textures in vdW materials, marking a major step toward low-dimensional topological spintronics. The emergence of these nontrivial magnetic configurations has been confirmed through various experimental techniques, such as Lorentz transmission electron microscopy (LTEM), magnetic force microscopy (MFM), and spin-polarized scanning tunneling microscopy (SP-STM). Among the rapidly growing family of 2D magnets, the ferromagnetic compound Fe$_3$GeTe$_2$ has attracted particular attention as a model system for hosting and manipulating chiral spin textures, owing to their distinct symmetry properties and tunable interlayer interactions.

\subsection{Ferromagnet Fe$_3$GeTe$_2$}

The layered vdW ferromagnet Fe$_3$GeTe$_2$ (FGT) has emerged as a model material for hosting magnetic skyrmions in two dimensions. Ding \textit{et al.}~\cite{ding2020observation} reported the first direct real-space observation of skyrmion bubbles in exfoliated FGT flakes using Lorentz-TEM and magnetic force microscopy. Their study revealed that although bulk FGT is centrosymmetric, chiral spin textures can be stabilized by the interplay of perpendicular anisotropy, dipolar interactions, and surface-induced inversion symmetry breaking that provides an effective DMI. This discovery established FGT as the first vdW magnet in which nontrivial topological textures were experimentally confirmed, opening the door for 2D skyrmionics.

\begin{figure}[H]
  \centering
  \includegraphics[width=\linewidth]{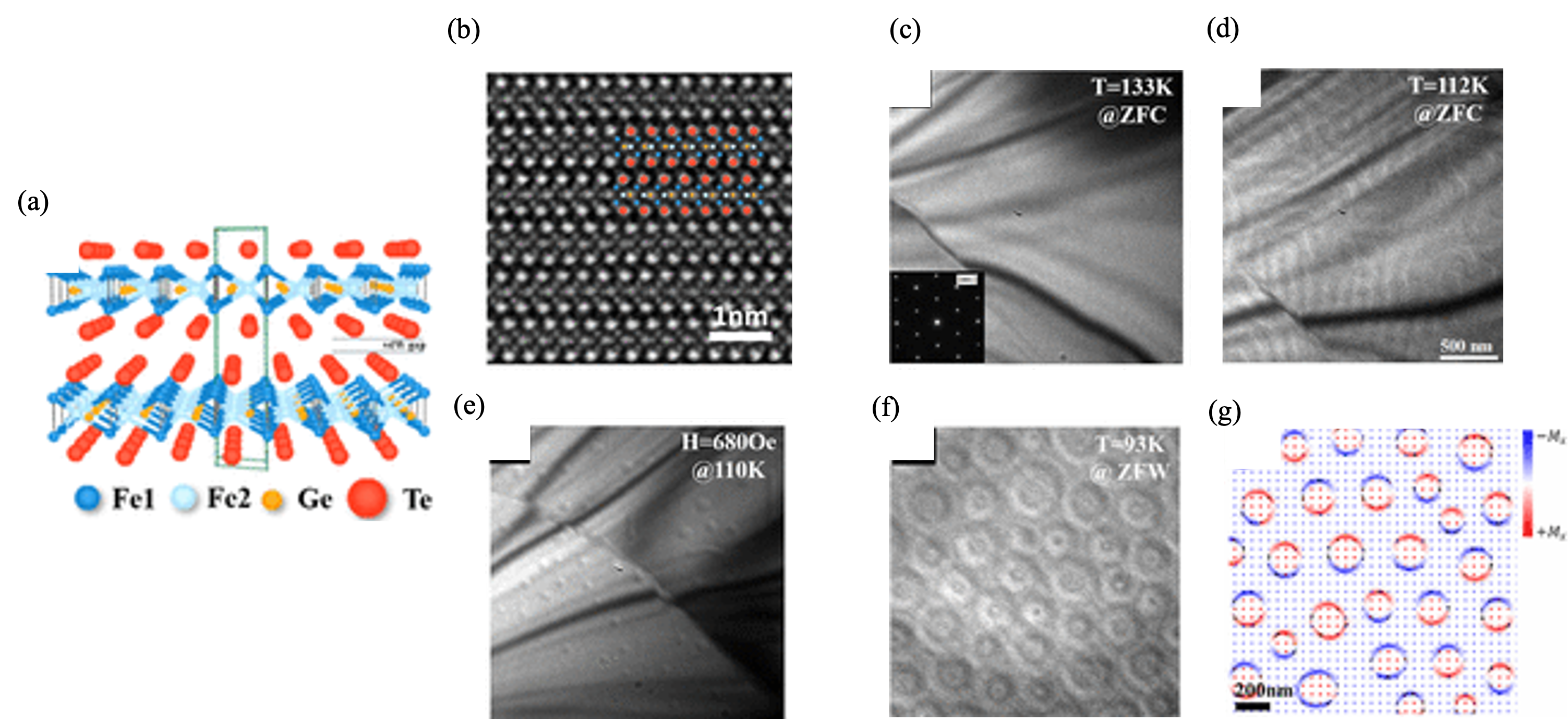}
  \caption{ (a) Schematic of the structure of a FGT bilayer with an interlayer vdW gap. (b) High-resolution STEM HAADF images along the [010] direction. (c)  The underfocused Lorentz-TEM images of FGT when the sample temperature is 133K. (d)  The underfocused Lorentz-TEM images of FGT when the sample temperature was lowered to 112K (a spontaneous ground state of the stripe domain). (e)  The underfocused Lorentz-TEM images showing magnetic-field-driven transitions from stripes gradually to bubbles. (f) The underfocused Lorentz-TEM images showing the bubble lattice evolution during a zero-field-warming (ZFW) process. (g) Theoretical simulation of skyrmion lattices at an applied magnetic field of 600~Oe for $\alpha = 0^\circ$. The scale bar is 200~nm.
Reproduced from \cite{ding2020observation}  with permission from the American Chemical Society. Copyright © 2020 ACS.}
  \label{fig:fig2}
\end{figure}

Figure~\ref{fig:fig2} summarizes the structural characteristics and magnetic-texture evolution in Fe$_3$GeTe$_2$, together with the theoretical interpretation of the associated magnetic phenomena. Panels (a) and (b) depict the FGT bilayer structure and a corresponding HAADF-STEM image along the [010] direction, confirming its van der Waals layered crystallography. Underfocused Lorentz-TEM imaging at 133\,K (panel c) reveals the appearance of stripe-like magnetic domains, which become the spontaneous zero-field ground state upon cooling to 112\,K (panel d) as magnetic anisotropy and dipolar interactions strengthen. The observed bright--dark contrast reflects Bloch-type domain walls consistent with FGT’s intrinsic ferromagnetism.
When a perpendicular magnetic field is applied (panel e), these stripes gradually transform into circular bubble-like magnetic textures, forming skyrmion-like domains stabilized not by intrinsic Dzyaloshinskii--Moriya interaction but by the interplay between strong out-of-plane anisotropy and magnetostatic energy. During zero-field warming (panel f), bubble lattices produced after field cooling remain metastable at low temperatures but progressively shrink and revert to stripes above approximately 134\,K.
Importantly, panel (g) presents a theoretical micromagnetic simulation of the skyrmion lattice under an applied magnetic field, illustrating the spatial distribution of the magnetization and reproducing the experimentally observed bubble-like textures. Together, experiment and theory demonstrate that FGT robustly hosts skyrmion-like bubble phases over a broad temperature window, driven predominantly by anisotropy-dipolar physics rather than chiral interactions.

\begin{figure}[htb!]
  \centering
  \includegraphics[width=\linewidth]{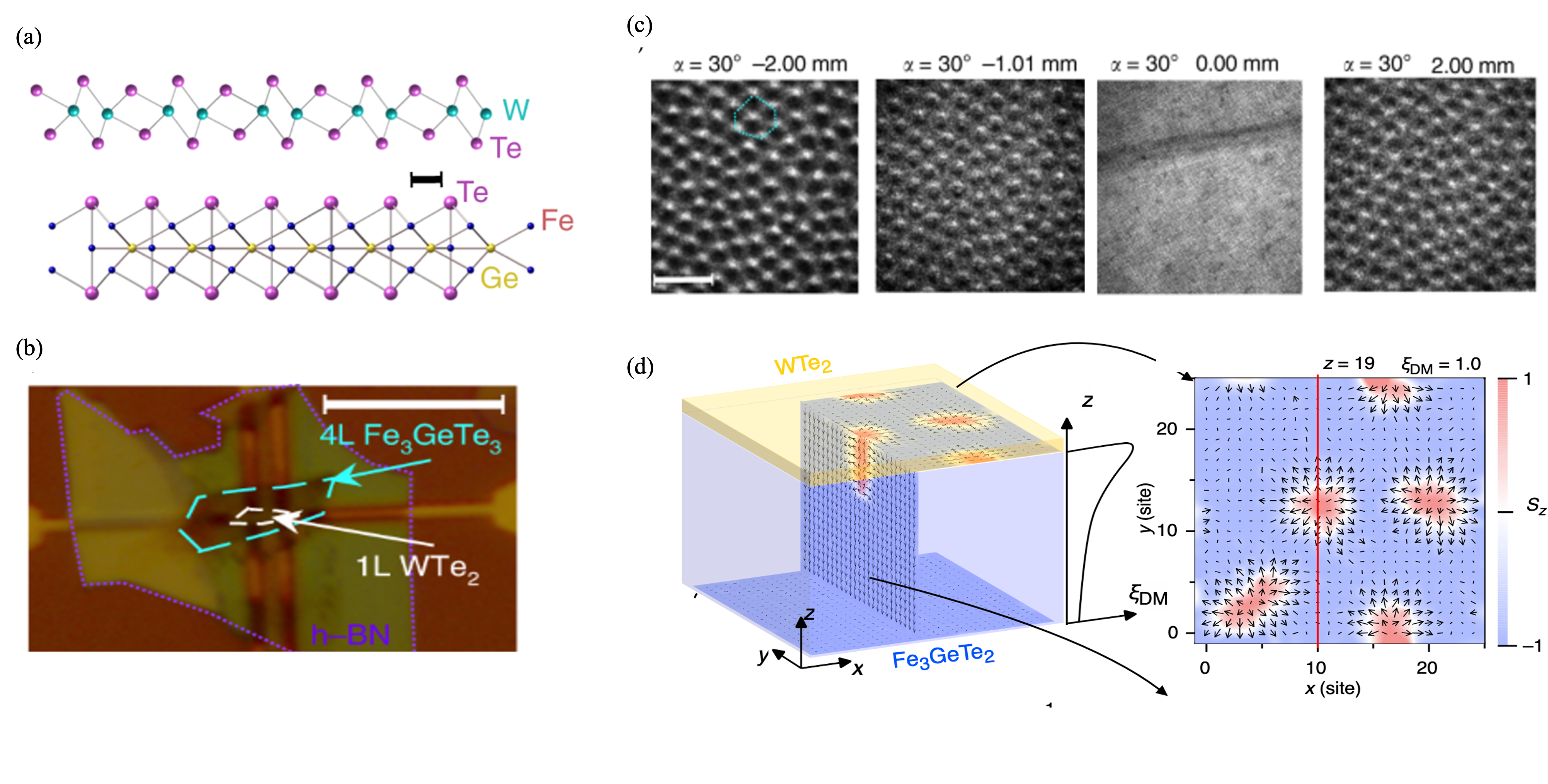}
\caption{N\'eel-type skyrmions in WTe$_2$/Fe$_3$GeTe$_2$ heterostructures (Wu \textit{et al.}, 2020). 
(a) Schematic crystal structure of monolayer WTe$_2$ stacked on Fe$_3$GeTe$_2$, illustrating the broken inversion symmetry at the interface. 
(b) Optical micrograph of a representative device composed of 1L WTe$_2$/4L Fe$_3$GeTe$_2$ encapsulated in h-BN (scale bar: 10\,$\mu$m). 
(c) Real-space Lorentz-TEM images of a skyrmion lattice in 1L WTe$_2$/40L Fe$_3$GeTe$_2$ at 180\,K under a perpendicular field of 510\,Oe, shown under a series of under-, in-, and over-focus conditions (scale bar: 500\,nm). 
(d) Micromagnetic and DFT-based simulation of the interfacial DMI, showing the localized antisymmetric exchange at the WTe$_2$/Fe$_3$GeTe$_2$ interface and the resulting N\'eel-type skyrmion spin texture.Reproduced under a Creative Commons Attribution 4.0 License ( CC BY)~\cite{Wu2020}.}
  \label{fig:wute}
\end{figure}

A complementary set of experiments by Wu \textit{et al.}~\cite{Wu2020} demonstrated that interfacial engineering can stabilize true N\'eel-type skyrmions in WTe$_2$/Fe$_3$GeTe$_2$ (FGT) heterostructures. By placing a monolayer of WTe$_2$, which is a semimetal with strong spin–orbit coupling, on top of few-layer FGT, inversion symmetry at the interface is broken, producing a sizable interfacial Dzyaloshinskii–Moriya interaction (DMI). This interfacial DMI fundamentally alters the topology of the spin textures compared to pristine FGT and enables the stabilization of robust N\'eel-type skyrmions with characteristic diameters of $\sim$150\,nm.
Figure~\ref{fig:wute}(a) illustrates the atomic stacking of the WTe$_2$/FGT heterostructure, highlighting the asymmetry responsible for the emergence of interfacial DMI. An optical micrograph of a representative device is shown in Figure~\ref{fig:wute}(b), where a monolayer of WTe$_2$ is assembled on top of multi-layer FGT and encapsulated in h-BN.
Direct real-space imaging of the magnetic texture is provided in Figure~\ref{fig:wute}(c), which displays Lorentz-TEM images acquired under a sequence of under-, in-, and over-focus conditions. The images reveal a dense, hexagonally arranged skyrmion lattice in the WTe$_2$/FGT heterostructure at 180\,K under an applied perpendicular field of 510\,Oe. The contrast reversal between bright and dark circular features across the different focus conditions reflects the in-plane curling of spins at the skyrmion boundary, which is a signature of N\'eel-type skyrmions. The persistence of this contrast and the well-defined periodicity confirm the stability and topological nature of the skyrmion lattice induced by the WTe$_2$ overlayer.
To clarify the microscopic origin of this behavior, Wu \textit{et al.} carried out first-principles  calculations and micromagnetic simulations, summarized in Figure~\ref{fig:wute}(d). The DFT calculations reveal that the interfacial DMI is strongly localized at the WTe$_2$/FGT boundary, where the combination of broken inversion symmetry and the large spin–orbit coupling of WTe$_2$ generates a substantial antisymmetric exchange interaction. The DMI magnitude decays rapidly with increasing distance from the interface, leading to a finite penetration depth into the FGT layer. Micromagnetic simulations using the DFT-derived DMI parameters reproduce the experimentally observed skyrmion size, lattice periodicity, and N\'eel-type spin configuration, confirming that the interfacial DMI is the dominant mechanism underlying the stabilized textures.

Beyond these two milestone works, subsequent studies have expanded the landscape of skyrmions in FGT. Transport experiments revealed a pronounced THE in exfoliated FGT, providing indirect evidence of chiral spin textures in addition to imaging~\cite{Wang2019THE}. Yang \textit{et al.}~\cite{Yang2020} demonstrated the emergence of N\'eel-type skyrmions in centrosymmetric vdW ferromagnet Fe$_3$GeTe$_2$ (FGT) through interfacial coupling with a perpendicularly magnetized [Co/Pd]$_n$ multilayer. By introducing a tunable Pd spacer layer, they controlled the strength of exchange interaction across the interface, which converted stripe domains in FGT into isolated skyrmionic bubbles even in the absence of an external magnetic field. The interfacial DMI  arising from broken inversion symmetry at the FGT/Pd interface was found to stabilize these N\'eel-type spin textures. This work highlights an effective strategy to engineer topological spin structures in centrosymmetric two-dimensional magnets via interlayer exchange coupling, offering a route to manipulate skyrmions through heterostructure design rather than intrinsic crystal asymmetry. Chemical substitution and stoichiometric tuning, for instance in Fe$_{3-x}$GaTe$_2$, stabilized robust N\'eel skyrmions at room temperature, even enabling optical writing and erasing of skyrmion states~\cite{Li2024}. These advances show that FGT and its derivatives can host a wide variety of skyrmionic states, ranging from metastable bubbles to stable, nanoscale N\'eel skyrmions, making them central players in the field of 2D skyrmionics.

Together, these works define the progression of skyrmion physics in vdW materials. 
In pristine FGT, skyrmion-like bubbles appear primarily through magnetostatic mechanisms and surface effects, existing as metastable states under controlled field–temperature histories. 
In contrast, in WTe$_2$/FGT heterostructures, engineered interfacial DMI stabilizes robust N\'eel skyrmions with well-defined chirality, detectable by both Lorentz-TEM and transport signatures. 
Later studies show that tuning interfaces~\cite{Wu_CGT_FGT2022}, coupling to multilayers~\cite{Yang2020}, controlling defects and vacancy distributions~\cite{Chakraborty2022}, local writing/erasing, or history-dependence and thickness control~\cite{Son2022} further expands the stability and functionality of these textures. 
This illustrates the design principle for 2D skyrmionics: while intrinsic vdW magnets provide a fertile platform for bubble-like topological textures, heterostructure engineering and material tuning allow deliberate control of symmetry and spin–orbit coupling to achieve stable, functional skyrmions suitable for devices.

\subsection{Ferromagnet Fe$_{3-x}$GaTe$_2$ and Fe$_3$GaTe$_2$}

\begin{figure}[H]
\centering
\includegraphics[width=\textwidth]{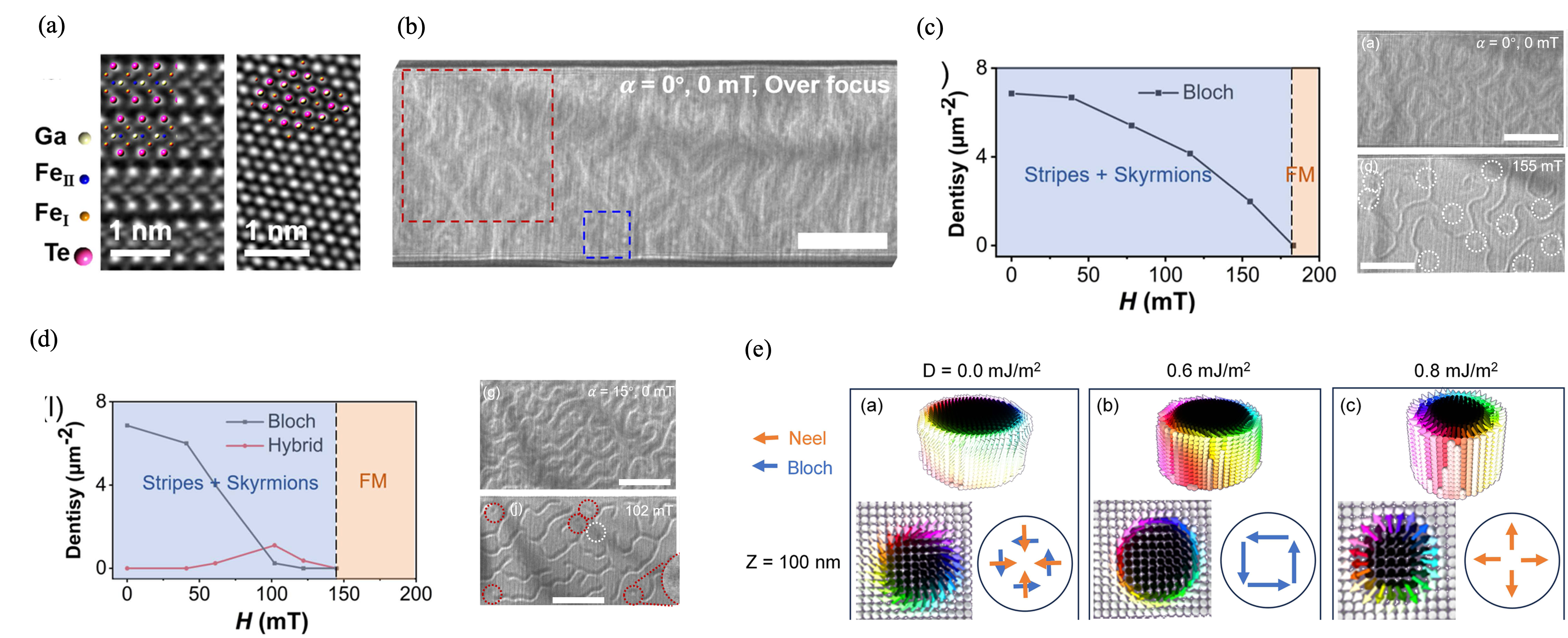}
\caption{
Topological spin textures in Fe$_3$GaTe$_2$ (Lv \textit{et al.}, 2024). 
(a) HAADF-STEM image along [001] resolving Ga, Fe$_\mathrm{I}$, Fe$_\mathrm{II}$, and Te columns. 
(b) Over-focused LTEM image at $\alpha = 0^\circ$, $H = 0$\,mT showing Bloch-type labyrinth domains. 
(c) Field-dependent Bloch skyrmion density with LTEM images showing the evolution from stripes (0\,mT) to coexisting stripes and Bloch skyrmions (155\,mT). 
(d) Under a tilted field ($\alpha = 15^\circ$), LTEM images reveal Bloch and hybrid skyrmions at 102\,mT, with hybrid density peaking near 100\,mT. 
(e) Micromagnetic simulations showing the transition from Bloch ($D = 0$) to hybrid ($D = 0.6$\,mJ\,m$^{-2}$) and N\'eel-type skyrmions ($D = 0.8$\,mJ\,m$^{-2}$). Reproduced under CC BY licence ~\cite{Lv2024_distinct}.}

\label{FeGaTe.png}
\end{figure}

In a recent study, Lv~\textit{et~al.}~\cite{Lv2024_distinct} explored the magnetic topology of the two-dimensional van der Waals ferromagnet Fe$_3$GaTe$_2$, revealing multiple coexisting skyrmion phases whose stability depends sensitively on crystal symmetry, magnetic field, and the presence of Fe-site vacancies. Atomic-resolution STEM imaging [Fig.~\ref{FeGaTe.png}(a)] resolves the Ga, Fe$_\mathrm{I}$, Fe$_\mathrm{II}$, and Te sublattices, confirming the high structural quality of Fe$_3$GaTe$_2$ and establishing a baseline for understanding the spin textures observed.
Using Lorentz transmission electron microscopy (LTEM), the authors showed that Fe$_3$GaTe$_2$ exhibits extended Bloch-type labyrinth domains at zero magnetic field, as seen in the over-focused LTEM image in Fig.~\ref{FeGaTe.png}(b). These Bloch-type stripe patterns reflect the centrosymmetric crystal symmetry, which favors Bloch walls in the absence of sizable DMI.
Upon increasing the magnetic field, the Bloch skyrmion density decreases monotonically [Fig.~\ref{FeGaTe.png}(c)], eventually vanishing near 183\,mT as the system transitions to the ferromagnetic (FM) state. The corresponding LTEM images show the evolution from labyrinth domains at 0\,mT to the coexistence of stripe domains and Bloch skyrmions at 155\,mT, indicating a continuous field-driven transformation of the domain topology.
A distinct behavior emerges when the sample is imaged under a tilted magnetic field ($\alpha = 15^\circ$). As shown in Fig.~\ref{FeGaTe.png}(d), Bloch and hybrid skyrmions coexist around 100\,mT, with the hybrid skyrmion population peaking before collapsing at higher fields. The hybrid skyrmions (red dashed circles) exhibit mixed Bloch–N\'eel character resulting from the combined action of dipolar interactions and a weak symmetry-allowed DMI. Their presence highlights the competition between Bloch-favoring dipolar energies and N\'eel-favoring interfacial DMI.
Theoretical micromagnetic simulations further support these observations, as illustrated in Figs.~\ref{FeGaTe.png}(e). For zero DMI ($D = 0$\,mJ\,m$^{-2}$), the simulations stabilize a pure Bloch-type skyrmion configuration. Introducing a moderate DMI of $D = 0.6$\,mJ\,m$^{-2}$ yields hybrid skyrmions with mixed Bloch and N\'eel components, consistent with tilted-field LTEM observations. At larger DMI ($D = 0.8$\,mJ\,m$^{-2}$), the skyrmion core evolves toward a N\'eel-type configuration, demonstrating the tunability of skyrmion helicity through DMI strength.
Together, the experimental and theoretical results establish Fe$_3$GaTe$_2$ as a versatile 2D magnetic system hosting Bloch, hybrid, and intermediate skyrmion configurations whose stability is governed by a delicate balance between dipolar interactions and emergent DMI.

\begin{figure}[H]
\centering
\includegraphics[width=\textwidth]{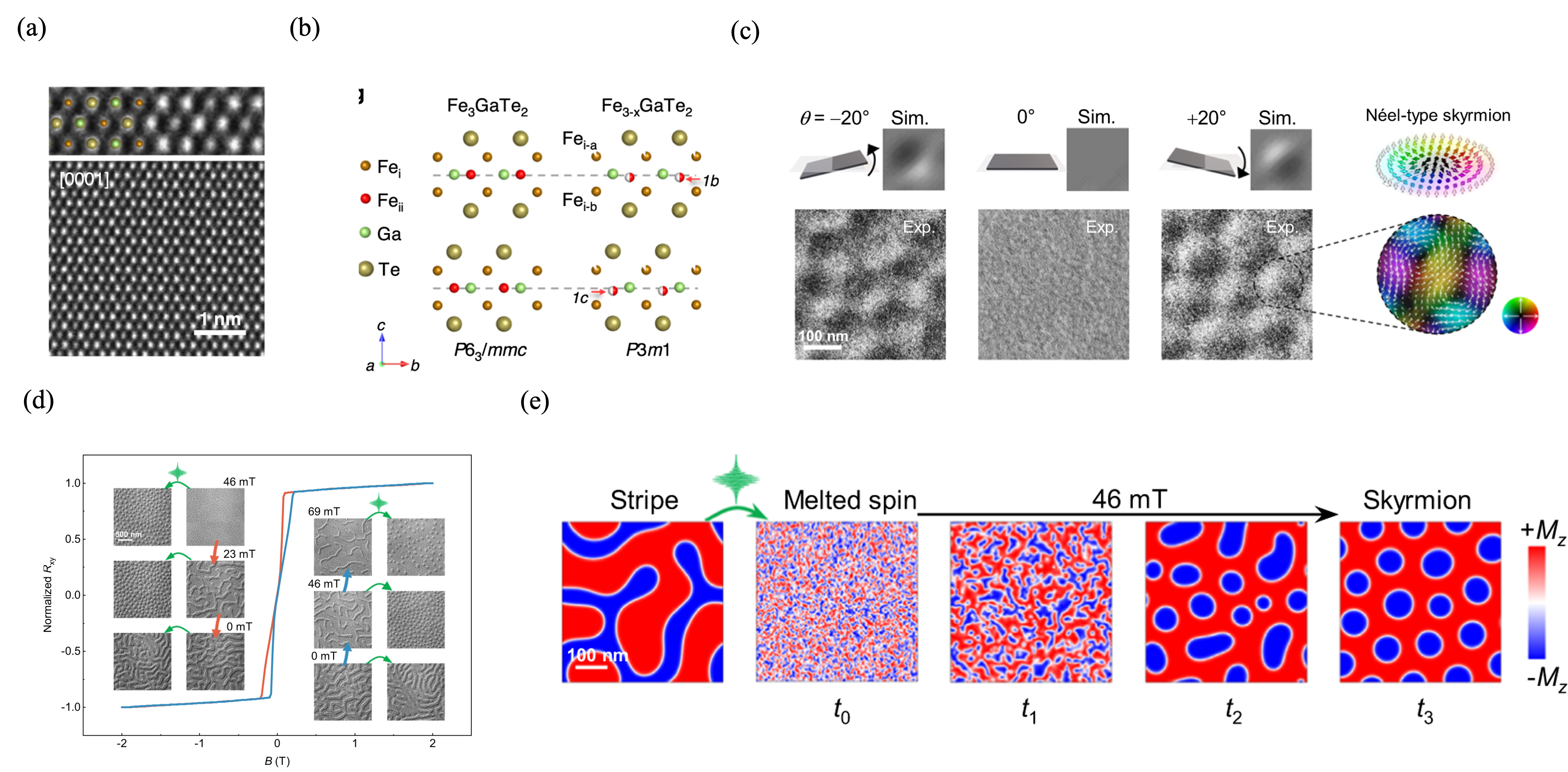}
\caption{
(a) HAADF-STEM image of Fe$_{3-x}$GaTe$_2$ along [0001], resolving Fe, Ga, and Te columns and revealing Fe-site deficiency. 
(b) Structural comparison of Fe$_3$GaTe$_2$ and Fe$_{3-x}$GaTe$_2$, showing Fe$_\mathrm{II}$ vacancies that break inversion symmetry and enable a finite DMI. 
(c) Experimental and simulated LTEM images at tilt angles $\theta = -20^\circ$, $0^\circ$, and $+20^\circ$, confirming N\'eel-type skyrmion contrast. 
(d) Anomalous Hall resistivity and LTEM snapshots showing the field-driven evolution from stripe domains to mixed stripe–skyrmion states. 
(e) Micromagnetic simulations of femtosecond-laser–induced domain melting and skyrmion formation under a 46\,mT field. Reproduced from Ref.~\cite{Li2024} under CC BY 4.0 license. }
\label{fig:lv_intro}
\end{figure}

Building on these findings, Lv~\textit{et~al.} further investigated Fe$_{3-x}$GaTe$_2$ to uncover the microscopic origin and control of topological spin textures~\cite{Li2024}. Atomic-scale HAADF-STEM measurements [Fig.~\ref{fig:lv_intro}(a)] reveal displaced Fe$_\mathrm{II}$ sites and local lattice distortions associated with Fe deficiency. A direct comparison between the ideal Fe$_3$GaTe$_2$ crystal structure and the Fe$_{3-x}$GaTe$_2$ model [Fig.~\ref{fig:lv_intro}(b)] shows that Fe-site vacancies lower the symmetry from centrosymmetric $P6_3/mmc$ to noncentrosymmetric $P3m1$, thereby breaking inversion symmetry and enabling a finite DMI.
The emergence of N\'eel-type skyrmions in this defect-induced asymmetric structure is confirmed by a combined LTEM and simulation analysis [Fig.~\ref{fig:lv_intro}(c)]. By tilting the sample across $\theta = -20^\circ$, $0^\circ$, and $+20^\circ$, the characteristic half-dark/half-bright contrast expected for N\'eel-type skyrmions becomes visible, in excellent agreement with simulated LTEM images and the reconstructed spin texture.
To examine the magnetic-field evolution of domain topology, anomalous Hall measurements were performed alongside LTEM imaging [Fig.~\ref{fig:lv_intro}(d)]. The LTEM snapshots show field-driven transitions from stripe domains to mixed stripe–skyrmion states, which correlate with distinct features in the Hall hysteresis curves, confirming the strong coupling between transport response and real-space spin textures.
Finally, the authors demonstrated optical control of skyrmion formation in Fe$_{3-x}$GaTe$_2$. Micromagnetic simulations reproduce the femtosecond-laser–induced transformation pathway [Fig.~\ref{fig:lv_intro}(e)]: a stripe domain state melts into a disordered spin configuration ($t_0$) and subsequently recrystallizes into isolated N\'eel-type skyrmions ($t_1$–$t_3$) under an applied field of 46\,mT. This ultrafast demagnetization and rapid quenching mechanism provides a robust route for optically writing skyrmions in a room-temperature 2D magnet.

Overall, these complementary studies reveal how controlled structural symmetry breaking in Fe$_{3-x}$GaTe$_2$ enables the emergence of DMI-driven N\'eel-type skyrmions, whereas the parent compound Fe$_3$GaTe$_2$ hosts dipolar- and DMI-balanced Bloch and hybrid skyrmions. Together, they provide a unified platform for understanding and engineering skyrmion chirality, stability, and optical control in vdW ferromagnets—opening promising avenues for low-power, multi-state spintronic memory and logic devices.

\subsection{Ferromagnetic (Fe$_{0.5}$Co$_{0.5}$)$_5$GeTe$_2$ (FCGT)}

\begin{figure}[H]
    \centering
    \includegraphics[width=\textwidth]{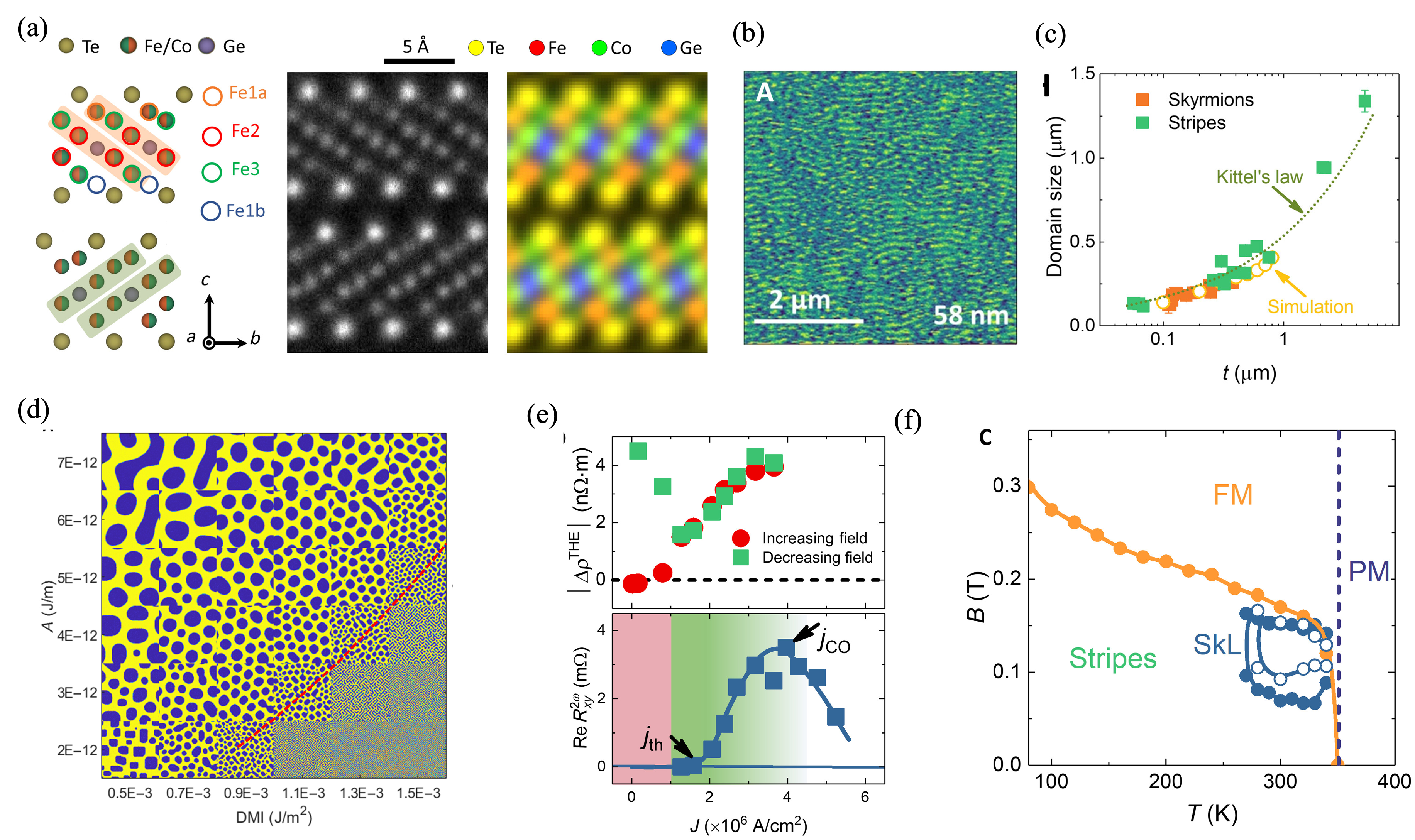}
    \caption{
Room-temperature N\'eel-type skyrmion lattice in (Fe$_{0.5}$Co$_{0.5}$)$_5$GeTe$_2$ (FCGT).
(a) Polar AA$'$-stacked structure breaking inversion symmetry, enabling in-plane DMI. 
(b) Lorentz-STEM induction map showing N\'eel-type spin rotation. 
(c) Thickness dependence of skyrmion size following Kittel’s law. 
(d) Micromagnetic phase map identifying the skyrmion stability region. 
(e) Topological Hall signal showing current-driven skyrmion motion above $1 \times 10^6$~A\,cm$^{-2}$. 
(f) Temperature--field phase diagram. Reproduced from Ref.~\cite{zhang2022sciadv} under(CC BY 4.0) licence.}
    \label{fig:fcgt_summary}
\end{figure}

Among the most notable recent advances in two-dimensional magnetism is the discovery of a 
room-temperature N\'eel-type skyrmion lattice in the layered polar magnet 
(Fe$_{0.5}$Co$_{0.5}$)$_5$GeTe$_2$ (FCGT), as reported by Zhang \textit{et al.}~\cite{zhang2022sciadv}. 
Figure~\ref{fig:fcgt_summary} summarizes the key structural, magnetic, and transport characteristics 
that underpin the stabilization of skyrmions in this material. The 
AA$'$-stacked crystal structure of FCGT (Fig.~\ref{fig:fcgt_summary}(a)) breaks inversion symmetry, 
giving rise to an uncompensated in-plane DMI, which serves as 
the fundamental mechanism stabilizing chiral N\'eel-type spin textures without requiring interfacial 
engineering. The Lorentz-STEM magnetic induction map (Fig.~\ref{fig:fcgt_summary}(b)) reveals the 
radial spin rotation pattern characteristic of a N\'eel-type configuration, providing direct experimental 
confirmation of the chiral nature of skyrmions at room temperature. The dependence of skyrmion formation on flake thickness follows Kittel’s law 
(\(d \propto t^{1/2}\)), where \(d\) denotes the skyrmion diameter and \(t\) the flake thickness, as shown in Fig.~\ref{fig:fcgt_summary}(c), indicating that magnetostatic and anisotropy energies jointly determine 
the equilibrium skyrmion size, with the most stable lattices observed for thicknesses between 
120~nm and 300~nm.
Complementary micromagnetic simulations (Fig.~\ref{fig:fcgt_summary}(d) delineate the stability 
region in parameter space, showing that skyrmion lattices emerge when the domain-wall energy nearly vanishes, i.e., when exchange, anisotropy, and DMI energies are delicately balanced. In addition to 
static stability, FCGT skyrmions exhibit current-driven dynamics with remarkable efficiency: 
the topological Hall resistivity (Fig.~\ref{fig:fcgt_summary}(e)) decreases above a threshold 
current density of approximately $1 \times 10^6$~A\,cm$^{-2}$, indicating the onset of skyrmion depinning 
and motion. Finally, the temperature--field phase diagram (Fig.~\ref{fig:fcgt_summary}(f)) 
summarizes the overall magnetic behavior, showing that the skyrmion lattice phase is stable between 
270~K and 340~K under moderate out-of-plane magnetic fields ($\sim$0.1~T). Collectively, these results 
establish FCGT as a prototype two-dimensional material hosting robust, tunable, and electrically 
controllable N\'eel-type skyrmions at room temperature, providing an important platform for advancing 
topological spin textures and low-power spintronic applications in vdW magnets.

\subsection{Chromium magnets Cr$_{1+\delta}$Te$_2$ and CrBr$_3$}

\begin{figure}[H]
    \centering
    \includegraphics[width=\textwidth]{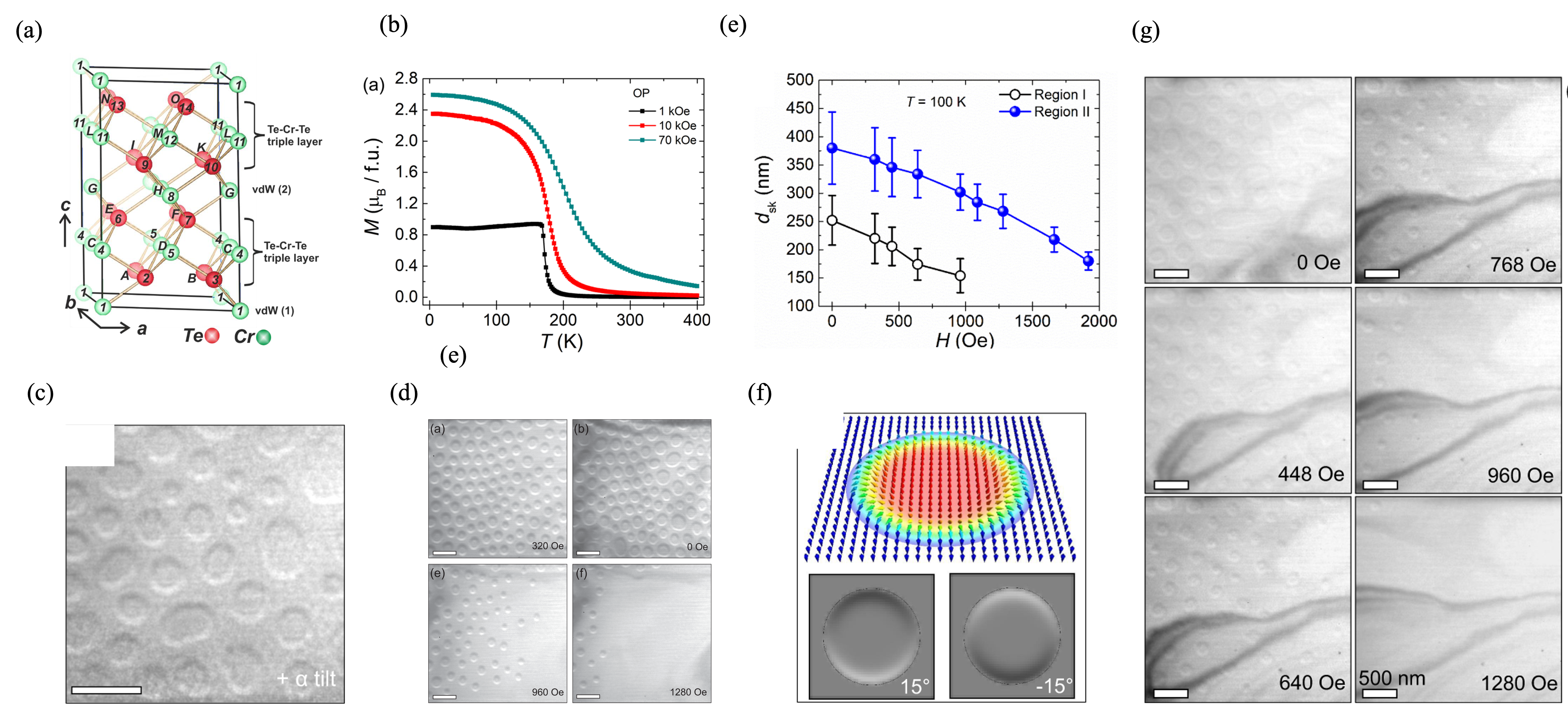}
\caption{
(a) Crystal structure of self-intercalated Cr$_{1+\delta}$Te$_2$, showing inequivalent vdW gaps that break inversion symmetry and yield a polar $P3m1$ structure.  
(b) Out-of-plane magnetization $M(T)$ showing strong perpendicular anisotropy and $T_C \approx 200$~K.  
(c) Skyrmion diameter $d_{\mathrm{sk}}$ vs.\ magnetic field at 100~K for two sample regions.  
(d) Low-field LTEM image showing circular N\'eel-type skyrmion contrast.  
(e) LTEM sequence (0–1280~Oe) showing skyrmion evolution and collapse with increasing field.  
(f) Simulated N\'eel-type skyrmion and corresponding LTEM contrast for $\pm 15^\circ$ tilts.  
(g) Tilted ($\alpha = +6^\circ$) LTEM images at various fields for a $\sim$75~nm lamella, confirming N\'eel-type contrast.  
Reproduced from~\cite{saha2022natcomm} under CC BY licence.}
    \label{fig:crte_summary}
\end{figure}

In the study of chiral magnetism in van der Waals (vdW) materials, chromium-based compounds provide a versatile platform not only for stabilizing skyrmions via intrinsic symmetry breaking, but also for controlling their chirality and collective behavior. In particular, self-intercalated Cr$_{1+\delta}$Te$_2$ and exfoliated CrBr$_3$ represent two complementary realizations of this physics, where the former establishes the microscopic origin of the Dzyaloshinskii--Moriya interaction (DMI), while the latter demonstrates dynamic chirality control under external fields.
Saha \textit{et al.}~\cite{saha2022natcomm} demonstrated that in acentric Cr$_{1+\delta}$Te$_2$, partial self-intercalation of Cr atoms within the vdW gaps breaks inversion symmetry and generates a bulk DMI, enabling the stabilization of chiral spin textures without the need for heterointerfaces. Figure~\ref{fig:crte_summary} summarizes the structural and magnetic signatures underpinning this behavior.
The atomic structure in Fig.~\ref{fig:crte_summary}(a) shows Te (red) and Cr (green) atoms arranged in the $y=0$ and $y=\frac{1}{2}$ planes. Two inequivalent vdW gaps, labeled (1) and (2), create distinct local coordination environments that drive vertical displacements of adjacent Te--Cr--Te trilayers. This structural asymmetry lowers the symmetry from centrosymmetric $Pm$ to polar $P3m1$, providing the inversion-symmetry breaking required for a finite bulk DMI.
Magnetization measurements in Fig.~\ref{fig:crte_summary}(b) reveal strong perpendicular magnetic anisotropy with a Curie temperature of approximately 200~K. This out-of-plane anisotropy stabilizes the skyrmion cores against thermal fluctuations. The field dependence of the skyrmion diameter in Fig.~\ref{fig:crte_summary}(c) shows a monotonic decrease of $d_{\mathrm{sk}}$ with increasing out-of-plane magnetic field. In addition, thicker lamellae host larger skyrmions, reflecting the enhanced role of dipolar interactions in thicker samples.
Direct real-space imaging using Lorentz transmission electron microscopy (LTEM) provides clear evidence of N\'eel-type skyrmions. In Fig.~\ref{fig:crte_summary}(d), circular magnetic contrast is observed at low magnetic fields, consistent with isolated N\'eel-type skyrmions. The field evolution shown in Fig.~\ref{fig:crte_summary}(e) demonstrates that the skyrmion lattice progressively shrinks and eventually collapses into a field-polarized ferromagnetic state as the magnetic field increases from 0 to 1280~Oe. The simulated spin texture and corresponding LTEM contrast in Fig.~\ref{fig:crte_summary}(f), obtained under $\pm 15^\circ$ sample tilts, reproduce the characteristic contrast expected for N\'eel-type skyrmions, providing theoretical confirmation of their topology.
Further confirmation is obtained from tilted-LTEM imaging in Fig.~\ref{fig:crte_summary}(g), where asymmetric contrast is observed for a specimen tilted by $\alpha = +6^\circ$ in a lamella of thickness $\sim 75$~nm. The contrast reversal with tilt direction is a hallmark of N\'eel-type helicity, confirming the chiral nature of the skyrmions stabilized in Cr$_{1+\delta}$Te$_2$.

While these results establish the microscopic origin and stabilization of skyrmions in Cr$_{1+\delta}$Te$_2$, they also raise the broader question of whether skyrmion chirality, once defined by intrinsic symmetry breaking, can be dynamically controlled in vdW magnets. 
Recent work by Fullerton \textit{et al.}~\cite{fullerton2025adma} demonstrates that such control is indeed possible in exfoliated CrBr$_3$. Using cryogenic LTEM, they showed that the application of an in-plane magnetic field modifies the magnetic free energy landscape, enabling stochastic switching between skyrmions of opposite chirality. This behavior originates from the competition between the intrinsic DMI and the Zeeman coupling to the external magnetic field, which biases the system toward a preferred handedness. As a result, chirality becomes a dynamic degree of freedom rather than a fixed property. A key consequence of this tunability is the emergence of collective phase transitions in the skyrmion ensemble. At low in-plane fields, the system exhibits a disordered skyrmion liquid characterized by strong fluctuations and weak positional correlations. As the field increases, chirality switching becomes suppressed, leading to a freezing transition into a hexatic phase and eventually a long-range ordered skyrmion crystal.
Together, these results establish a unified picture of chiral magnetism in vdW chromium compounds. In Cr$_{1+\delta}$Te$_2$, chemical self-intercalation induces polar symmetry and a sizable intrinsic DMI, enabling the formation of robust N\'eel-type skyrmions in a bulk material. In contrast, CrBr$_3$ demonstrates that skyrmion chirality can be dynamically tuned by external fields, providing a powerful route to control both the topology and collective phases of skyrmions. This combination of intrinsic DMI engineering and external-field control opens new pathways for designing tunable chiral spin textures in low-dimensional magnetic systems.

\section{Theoretical studies of topological magnetism in 2D vdW materials}

Certainly, theoretical studies and predictions have played a crucial role in identifying vdW materials as promising candidates for the emergence of topological magnetic states, while also providing insight into the underlying mechanisms.
Following the overview of experimental studies, we now discuss several selected theoretical highlights.

As mentioned earlier, Fe$_3$GeTe$_2$ has emerged as a prototypical itinerant two-dimensional ferromagnet, yet theoretical studies did not initially predict topological spin textures in this system. Early first-principles investigations focused on its magnetic anisotropy and exchange interactions. For example, Zhuang \textit{et al.}~\cite{Zhuang2016FGT} used DFT to demonstrate strong perpendicular magnetic anisotropy in monolayer FGT, establishing the microscopic ingredients necessary for stabilizing noncollinear states.
Following the experimental observation of skyrmion bubbles in FGT~\cite{{ding2020observation}}, theoretical work primarily played an explanatory role. Micromagnetic simulations performed in the experimental study reproduced the observed bubble-like textures and showed that they arise from the competition between dipolar interactions and perpendicular anisotropy, rather than intrinsic DMI. This indicates that, for pristine FGT, experiment preceded theory in identifying topological textures.
A transition toward predictive theoretical modeling occurred when heterostructures were considered. Zhang \textit{et al.}~\cite{Wu2020NatCommun} combined DFT with Monte Carlo simulations to show that interfacial symmetry breaking in WTe$_2$/FGT induces sizable DMI and stabilizes N\'eel-type skyrmions. Similarly, Xu \textit{et al.}~\cite{Xu2022FGTSk} used DFT and Monte Carlo methods to predict diverse skyrmionic phases in FGT-based systems under symmetry breaking and interaction tuning. These studies demonstrate that while theory did not originally predict skyrmions in pristine FGT, it has become a powerful tool for engineering chiral magnetism in FGT-based heterostructures.

In Fe$_3$GaTe$_2$ and Fe$_{3-x}$GaTe$_2$, theoretical insights are closely intertwined with experimental discoveries. Lv \textit{et al.}~\cite{Lv2024_distinct,Li2024} used micromagnetic simulations to show that Bloch-type skyrmions emerge in centrosymmetric Fe$_3$GaTe$_2$, while increasing DMI leads to hybrid and N\'eel-like textures. These simulations were essential for interpreting Lorentz-TEM observations and identifying the nature of the spin textures.
In Fe$_{3-x}$GaTe$_2$, Lv \textit{et al.}~\cite{Li2024} combined DFT and micromagnetic simulations to demonstrate that Fe deficiency lowers crystal symmetry and induces finite DMI, enabling room-temperature N\'eel-type skyrmions. In this material family, theory did not clearly precede experiment. Instead, DFT and micromagnetic modeling were developed concurrently with experimental work to identify the microscopic mechanisms, particularly symmetry breaking and disorder, that stabilize topological spin textures.

Theoretical studies of (Fe$_{0.5}$Co$_{0.5}$)$_5$GeTe$_2$ (FCGT) highlight the role of crystal symmetry in enabling intrinsic chiral magnetism. Based on first-principles, Zhang \textit{et al.}~\cite{Zhang2022PolarMetal} showed that Co substitution stabilizes a polar AA$'$ structure that breaks inversion symmetry and generates intrinsic DMI.
This theoretical insight emerged simultaneously with the experimental observation of room-temperature N\'eel-type skyrmion lattices~\cite{zhang2022sciadv}, indicating a parallel development of theory and experiment. Subsequent work by Li \textit{et al.}~\cite{Li2024Fe5GeTe2} employed DFT-derived spin Hamiltonians and atomistic simulations to predict stable nanoscale skyrmions in the Fe$_5$GeTe$_2$ family, including estimates of skyrmion size and lifetime.
Thus, in FCGT, theory and experiment progressed together, with theory providing a microscopic explanation for the symmetry-breaking mechanism and later evolving toward predictive modeling of skyrmion stability.

For acentric self-intercalated Cr$_{1+\delta}$Te$_2$, the experimental observation of N\'eel-type skyrmions~\cite{saha2022natcomm} preceded a direct theoretical prediction for the same material. However, earlier theoretical studies on related systems had already established the mechanisms necessary for stabilizing chiral magnetism.
Fragkos \textit{et al.}~\cite{Fragkos2022CrTe2WTe2} used first-principles calculations and spin modeling to show that CrTe$_2$/WTe$_2$ heterostructures host interfacial DMI and support skyrmions. 
%Abuawwad \textit{et al.}~\cite{Abuawwad2023PRB}  predicted a rich set of topological textures, including skyrmions, merons, and antimerons, in CrTe$_2$-based heterobilayers.
Further theoretical work demonstrated that CrTe$_2$-based systems host an even richer class of topological states beyond conventional skyrmions. In particular, Abuawwad \textit{et al.}~\cite{Abuawwad2023PRB} predicted the emergence of AFM multi-meronic states in CrTe$_2$ and its heterobilayers. These textures arise from a frustrated magnetic background that can be decomposed into three interpenetrating sublattices with 120$^\circ$ spin orientation. Within this framework, each sublattice hosts pairs of merons and antimerons, which are topological objects carrying fractional charges $q = \pm \frac{1}{2}$ depending on their vorticity and core polarization. The resulting magnetic state is a composite multi-meronic (hexamer) texture, where the total topological charge $Q$ emerges from the combination of sublattice contributions, allowing for configurations with $Q = 0$ as well as finite charges $Q = \pm 1$. This decomposition highlights a fundamentally different topological landscape compared to ferromagnetic skyrmions, where topology is encoded in a single continuous field, whereas here it is distributed across coupled AFM sublattices. 
Further first-principles studies demonstrated that external electric fields can directly modify the van der Waals gap and, consequently, the magnetic interactions in CrTe$_2$-based heterobilayers~\cite{Abuawwad2024}. In CrTe$_2$/RhTe$_2$, this enables switching between ferromagnetic skyrmions and meron pairs, while in CrTe$_2$/TiTe$_2$ the electric field enhances the stability of frustrated antiferromagnetic meronic states. This electrically driven control of competing interactions provides a mechanism to selectively stabilize distinct topological solitons within a single platform, highlighting the potential of vdW heterostructures for energy-efficient, electrically programmable spintronic devices. Therefore, while experiment came first for Cr$_{1+\delta}$Te$_2$, theory in the broader CrTe$_2$ family has been highly predictive and has significantly expanded the landscape of topological magnetism beyond experimentally observed phases.
Across these materials, different theory--experiment relationships emerge. In Fe$_3$GeTe$_2$, experiment preceded theory, which later provided explanations and predictive design strategies. In Fe$_3$GaTe$_2$ systems, theory and experiment were developed concurrently. In FCGT, both progressed in parallel, with theory explaining the polar structure responsible for intrinsic DMI. In contrast, the CrTe$_2$ family demonstrates a case where theory predicts a much richer set of skyrmions phases, which include merons, antimerons, and electrically controllable topological states, extending beyond current experimental realizations.

Strain, lattice distortion, and chemical asymmetry provide powerful microscopic routes for generating intrinsic DMI and frustrated exchange interactions in two-dimensional vdW magnets~\cite{Huang2024PRB}. These mechanisms have been systematically explored using first-principles calculations combined with atomistic spin modeling, enabling the extraction of magnetic interactions and the mapping of resulting topological phase diagrams. In particular, broken inversion symmetry—arising from lattice distortions, layer stacking, or chemical asymmetry—induces finite DMI, while competing exchange interactions promote magnetic frustration and the stabilization of noncollinear spin textures. A prominent realization of these principles is found in Janus van der Waals magnets, where mirror symmetry is intrinsically broken by chemical substitution between the two sides of the layer, leading to intrinsic chiral magnetic states, including polar skyrmions, as demonstrated for CrInX$_3$ (X = Se, Te)~\cite{Zhou2024JanusSkyrmion}. Complementarily, engineering inversion asymmetry through layer stacking provides an additional route to tailor chiral interactions and stabilize topological spin textures such as antiskyrmions in two-dimensional magnets~\cite{PhysRevB.109.024426}. Together, these results highlight that structural asymmetry, whether intrinsic or engineered, offers a robust pathway for realizing and controlling topological magnetism in low-dimensional materials.

Merons and antimerons emerge in systems with interlayer coupling or domain-wall asymmetry, forming meron–antimeron lattices under moderate fields. Easy-plane 2D magnets with weak perpendicular anisotropy instead host bimerons, topological analogues of skyrmions with in-plane cores, predicted in Fe$_5$GeTe$_2$ and VSe$_2$ where strain and doping tune anisotropy continuously between easy-axis and easy-plane regimes~\cite{Cheng2024AdvFunctMater}. These diverse states illustrate how the delicate balance among exchange, anisotropy, and DMI defines the topological phase diagram in two dimensions.
A 2D material such h-BN can be used to trigger structural transition, coined Kagometization, in a 3d transition metallic film deposited on a hexagonal heavy metallic surface. N atoms, due to their strong electronegativity, move 3d atoms to form a Kagome lattice, which can host a rich set of skymrionic or meronic textures~\cite{Zhou2024NatCommun}. For antiferromagnets, strong magnetic frustration drives multi-$q$ states, hosting large topological charges due to their non-coplanarity. The intrinsic frustration imposes  self-induced spin-glass-like behavior~\cite{Zhou2025npj}. Note that strained 2D materials such as h-BN and graphene deposited on magnetic films can modify the DMI and MAE even without structural transitions~\cite{Yang2017,Hallal2021}.

Frustrated and multi-$q$ magnets add further complexity. Competing exchange interactions in triangular or kagome lattices can produce vortex–antivortex crystals and multi-$q$ skyrmion lattices with noninteger topological charge. Monte-Carlo simulation based on parameters from DFT on MnSeTe and Fe/graphene predict coexisting skyrmion and antiskyrmion phases, stabilized by frustration and finite-temperature entropy~\cite{Arya2025JanusMnSeTe, Xu2020FeGraphene}. Such coexistence of multiple topological species expands the theoretical phase space from a single soliton solution to a correlated ensemble, suggesting that 2D materials may exhibit topological glasses or liquids composed of interacting quasiparticles.

A particularly promising theoretical development involves moiré-engineered topological lattices. Twisted bilayers of NiI$_2$, CrCl$_3$, or FePS$_3$ produce long-wavelength modulations of exchange and DMI that act as periodic potentials for spin textures. Analytical models predict “moiré skyrmions” whose helicity and periodicity depend on twist angle, bias, and interlayer separation~\cite{Akram2021MoireSkyrmions, akram2020twisted, Antao2024NiI2MoireSkyrmion}. Such artificial superlattices host programmable arrays of topological excitations whose lattice constants can exceed 100~nm, enabling direct optical or magneto-optical imaging. Theoretical calculations also reveal emergent dual-chirality domains—regions where N\'eel and Bloch skyrmions coexist within a single moir\'e period—illustrating how moir\'e physics provides a powerful symmetry-control mechanism for spin topology.

Defects, strain, and confinement play decisive roles in stabilizing or pinning these states. Atomistic modeling shows that atomic steps and vacancies modify local DMI vectors, serving as preferred nucleation sites for skyrmions or merons. Edge confinement generates potential wells that trap topological textures, enhancing their thermal stability and enabling guided motion along nanoribbons. Patterned strain gradients and electrostatic potentials have been proposed theoretically to control skyrmion trajectories and interactions, forming the basis for 2D magnetic racetrack devices. These results underscore how topology in vdW magnets is not a fixed property but a tunable and locally programmable degree of freedom~\cite{Chakraborty2022, Saha2024FGaT_DefectsSkyrmions, Sun2025FGT_GeometryStrain}.

Beyond static stability, theoretical efforts have increasingly focused on the dynamics and emergent electrodynamics of topological textures in two-dimensional van der Waals magnets. In this context, first-principles studies have demonstrated that topological solitons in 2D magnets can exhibit intrinsic magnetoelectric coupling, whereby noncollinear spin textures generate an electric polarization and can be controlled by external electric fields~\cite{Edstrom2025}. This finding elevates topological textures from purely magnetic quasiparticles to active multiferroic entities with coupled spin and charge degrees of freedom. 
In metallic vdW systems hosting skyrmion lattices, itinerant electrons that adiabatically follow the noncollinear spin background acquire a real-space Berry phase, which can be interpreted as an emergent magnetic field acting on charge carriers and giving rise to a topological Hall effect. Continuum and tight-binding models further predict that, for sub-10~nm skyrmions, these emergent fields can reach magnitudes of several tens of tesla, resulting in pronounced Hall responses in metallic two-dimensional platforms. Recent theoretical and experimental work on CrBr$_3$ further highlights that skyrmion chirality can be dynamically tuned by external magnetic fields, enabling switching between topological states and modifying the underlying energy landscape~\cite{fullerton2025adma}. Time-dependent spin-dynamics simulations further demonstrate current- and magnon-driven skyrmion motion in vdW magnets, characterized by high mobility and reduced Hall angles owing to low intrinsic damping. In parallel, theoretical proposals have shown that electric fields, interfacial polarization, or optical excitation can induce dynamic interconversion between skyrmions, bimerons, and antiskyrmions in vdW heterostructures, pointing toward reconfigurable topological circuits based on programmable spin textures
\cite{Fragkos2022CrTe2WTe2_THE,Zhang2020MagnonFGT,Sun2020BimeronVDW,Bo2025SkyrmionBimeronVDW}.

Higher-order magnetic interactions have recently emerged as a key ingredient in the theoretical description of topological magnetism in 2D vdW   materials~\cite{Kartsev2020Biquadratic,qkqn-lfzk}. While conventional models based on bilinear Heisenberg exchange and DMI capture chiral spin textures in non-centrosymmetric systems, they are often insufficient for atomically thin magnets, where reduced dimensionality, strong spin--orbit coupling, and multi-orbital effects enhance non-Heisenberg contributions. In this context, biquadratic and multi-spin interactions can significantly reshape the magnetic energy landscape, modify effective anisotropies, and compete with or complement DMI. These higher-order terms provide an alternative mechanism for stabilizing topological spin textures such as skyrmions, antiskyrmions, and merons, particularly in centrosymmetric or weak-DMI systems, by favoring noncoplanar configurations with finite scalar spin chirality. First-principles and atomistic studies further demonstrate that such interactions can control the size, shape, and stability of skyrmionic textures and enable complex topological phases in layered magnets such as Fe$_3$GeTe$_2$~\cite{qkqn-lfzk}, highlighting their central role in determining the magnetic phase diagram of two-dimensional vdW materials.

Recent studies have extended the investigation of topological magnetic textures in vdW systems toward externally driven and tunable regimes. In particular, first-principles and model-based works have demonstrated that electric fields and ferroelectric polarization provide an efficient route to control key magnetic interactions, such as magnetic anisotropy and interfacial DMI in vdW heterostructures. This enables reversible tuning of skyrmion stability, size, and chirality, as well as electrically driven creation and manipulation of skyrmionic states, highlighting the strong coupling between ferroelectric and magnetic degrees of freedom in these systems~\cite{Huang2022,Yao2023}. Such mechanisms emphasize the role of interfacial engineering and symmetry breaking in tailoring chiral spin textures in two-dimensional materials. In parallel, ultrafast laser-driven approaches have emerged as a powerful tool to access non-equilibrium regimes of topological magnetism in vdW magnets. Time-dependent simulations and experiments show that femtosecond optical excitation can induce rapid reconfiguration of spin textures, including the generation and switching of topologically nontrivial states. These processes are mediated by transient modifications of exchange interactions and magnetic anisotropy, enabling the stabilization of metastable spin configurations on ultrafast timescales~\cite{Strungaru2022,Khela2023}. Such light-induced control provides a pathway toward manipulating spin textures beyond equilibrium limits, offering access to transient magnetic phases that are otherwise inaccessible under static conditions.

A unifying insight emerging from the different theoretical studies is the central role of symmetry breaking, magnetic frustration, and interfacial engineering in generating chiral interactions and stabilizing nontrivial spin textures. In addition, external control parameters such as electric fields, strain, stacking geometry, and optical excitation provide versatile means to tune magnetic interactions and dynamically access non-equilibrium topological states. Together, these advances establish vdW materials as a highly tunable platform in which topology is not an intrinsic fixed property, but rather an emergent and controllable degree of freedom. This positions two-dimensional magnets at the forefront of efforts to realize programmable, energy-efficient, and potentially quantum-enabled spin-topological functionalities.

\section{Outlook and Challenges}

The field of topological magnetism in 2D vdW materials and heterostructures stands at the frontier of condensed matter physics, merging fundamental spin–orbit phenomena with emergent spin textures such as skyrmions, merons, and domain-wall solitons. Despite rapid progress since the discovery of intrinsic ferromagnetism in monolayers such as CrI$_3$, CrGeTe$_3$, and Fe$_3$GeTe$_2$, major theoretical and experimental challenges remain before these systems can deliver robust, room-temperature, and scalable functionalities.

From the theoretical side, key challenges lie in extending current modeling frameworks beyond collinear magnetism and standard density-functional theory. As we know, noncollinear and chiral configurations—driven by the DMI, frustrated exchange, or dipolar fields—require multiscale modeling that bridges ab initio and micromagnetic simulations. The finite-temperature dynamics of skyrmions and merons, their first-principles interactions with defects~\cite{LimaFernandes2018, LimaFernandes2020, Arjana2020,Reichhardt2022}, and their coupling to magnons, phonons, and conduction electrons remain largely unresolved. Existing theoretical descriptions often neglect these couplings, which are critical for predicting the real-time stability, diffusion, and annihilation pathways of topological spin textures. Furthermore, developing quantitative models for interlayer DMI, higher-order magnetic interactions, and spin–orbit torque (SOT) in vdW heterostructures remains an open frontier for both theory and simulation.

Experimentally, challenges are equally formidable. The weak DMI in pristine monolayer magnets limits the stabilization of chiral textures, motivating the exploration of broken-symmetry systems such as Janus transition-metal dichalcogenides (e.g., MnSeTe, CrSTe) and proximity-engineered heterostructures (e.g., CrI$_3$/WSe$_2$, Fe$_3$GeTe$_2$/Pt). However, interface quality and atomic registry play decisive roles: even small variations in stacking angle or interlayer distance can change the sign and magnitude of DMI, altering the topological ground state. Imaging magnetic textures in monolayers represents another challenge. While Lorentz TEM, spin-polarized STM, and NV magnetometry have succeeded in thicker films, few techniques can resolve skyrmions or merons at the single-nanometer scale in 2D vdW materials. Improved optical and scanning-probe methods are essential for directly correlating local spin textures with transport signatures such as the THE.

Looking ahead, opportunities abound in functional design and applications. The integration of 2D magnets with ferroelectric or topological layers could enable voltage-controlled magnetoelectric skyrmions, paving the way for low-power spintronic and neuromorphic computing. Moiré superlattices and twisted bilayer magnets open an entirely new regime, where exchange modulation induces moiré skyrmion lattices and tunable chiral spin networks. In metallic vdW systems like Fe$_3$GeTe$_2$ and CrTe$_2$, spin–orbit torque and AHE offer avenues for efficient current-driven manipulation of topological states. Furthermore, coupling magnetic textures to Dirac or Weyl fermions in topological insulators may yield exotic hybrid excitations, such as emergent magnetic monopoles or, when superconductivity is induced via the proximity effect, Majorana modes. The latter are promising candidates as building blocks for quantum computation. 

In summary, realizing and controlling topological magnetism in 2D vdW materials requires progress across multiple fronts: enhanced interfacial DMI, quantitative finite-temperature modeling, advanced imaging and characterization, and materials-by-design synthesis. Meeting these challenges will not only deepen our understanding of low-dimensional spin textures but also enable novel spintronic and quantum technologies that exploit the topological robustness and nanoscale tunability of 2D magnetic systems.

\section*{Acknowledgments}

N.A acknowledges funding from the European High Performance Computing Joint Undertaking (EuroHPC JU) under the project NA MaX – Materials Design at the eXascale, grant agreement No. 101093374. S.L. acknowledges funding by Deutsche Forschungsgemeinschaft (DFG): project 328545488 – CRC/
TRR 227, Project No. B12 and LO 1659/10-1 as well as the German Excellence Strategy –EXC3112/1 –533767171 (Center for Chiral Electronics).

\section*{References}
\bibliographystyle{elsarticle-num}
\bibliography{references} 

\end{document}